\documentclass{cernyrep}
\include{epstopdf}
\usepackage{cite}
\graphicspath{{//cern.ch/dfs/Users/r/rudi/Public/figures/JASpapers/}}
\usepackage{varwidth}
\usepackage{xcolor}

\begin{document}

\title{Machine Protection and Interlock Systems for Circular Machines---Example for LHC}
\author{R.~Schmidt}
\institute{CERN, Geneva, Switzerland}
\maketitle

\begin{abstract}
This paper introduces the protection of circular particle accelerators from accidental beam losses. Already the energy stored in the beams for accelerators such as the TEVATRON at Fermilab and Super Proton Synchrotron (SPS) at CERN could cause serious damage in case of uncontrolled beam loss. With the CERN Large Hadron Collider (LHC), the energy stored in particle beams has reached a value two orders of magnitude above previous accelerators and poses new threats with respect to hazards from the energy stored in the particle beams. A single accident damaging vital parts of the accelerator could interrupt operation for years. Protection of equipment from beam accidents is mandatory. Designing a machine protection system requires an excellent understanding of accelerator physics and operation to anticipate possible failures that could lead to damage. Machine protection includes beam and equipment monitoring, a system to safely stop beam operation (e.g. extraction of the beam towards a dedicated beam dump block or stopping the beam at low energy) and an interlock system providing the glue between these systems. This lecture will provide an overview of the design of protection systems for accelerators and introduce various protection systems. The principles are illustrated with examples from LHC.
\end{abstract}

\section{Designing a protection system for particle accelerators}

The approach to the design of a machine protection system (MPS) starts with the identification of the hazards. A number of failures are identified that can change beam parameters and lead to the loss of particles that would hit the aperture and possibly damage equipment. A simple procedure to get started includes several steps.

\begin{enumerate}
\item There is a very large number of failures that can cause beam losses. The failures are classified in different categories.
\item The risk for each failure (or for categories of failures) is estimated.
\item The worst-case failures and their consequences are identified.
\item Prevention of a failure or mitigation of the consequences of a failure is worked out.
\item This allows to start with the design of systems for machine protection.
\item Back to item 1.
\end{enumerate}

The design of the systems for machine protection starts during the early design phase of an accelerator. Since new hazards are identified during the life cycle of an accelerator, in particular during upgrades and modifications, new mitigation methods are developed; the process continues throughout the life cycle. It is required until end of operation.

\section{Circular accelerators and LHC}

\begin{figure}
\centering
\includegraphics[width=0.5\linewidth]{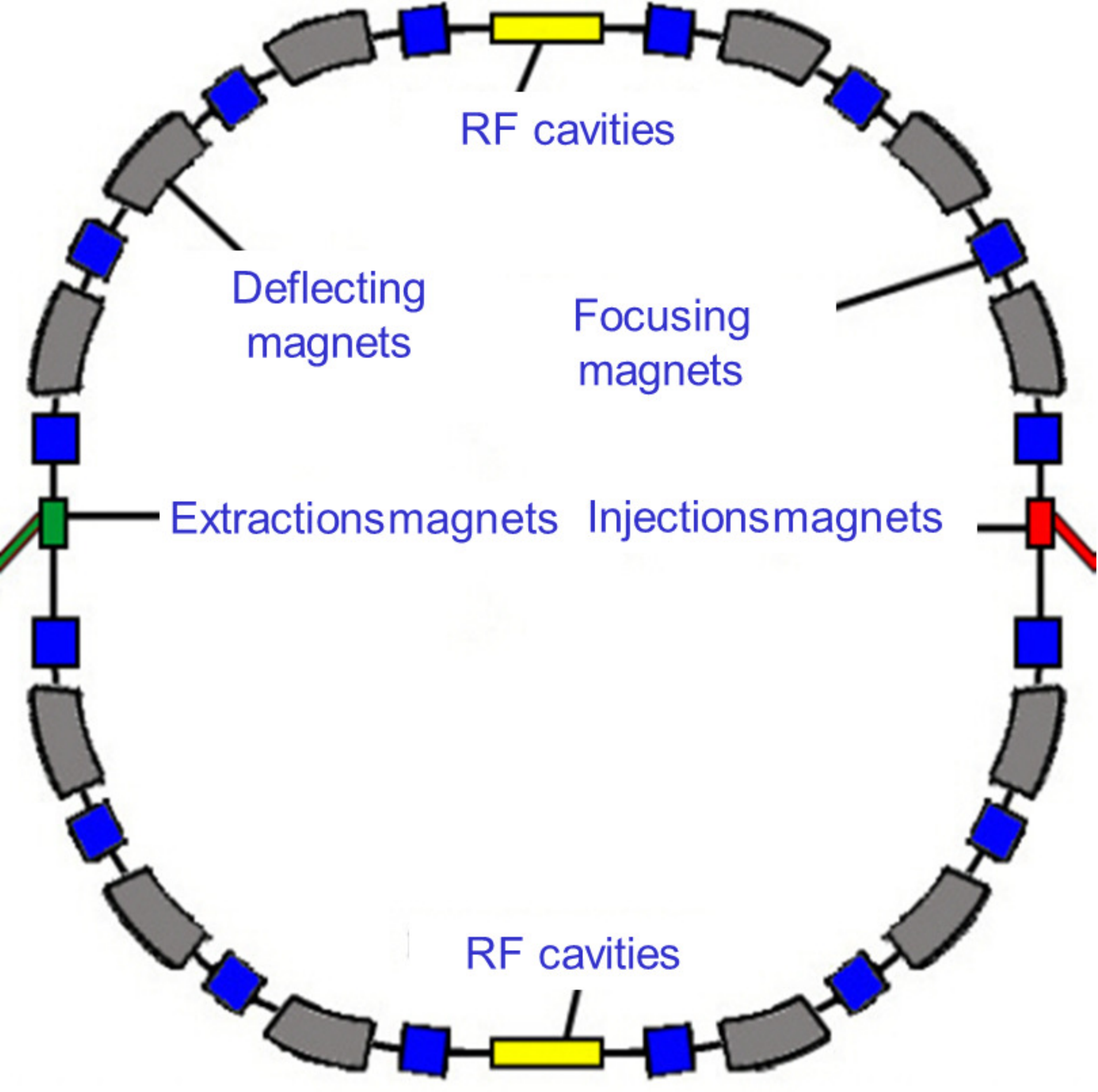}
\caption{Typical components of a circular accelerator}
\label{Components-Circular-Accelerator}
\end{figure}

The main components of a synchrotron are deflecting magnets, magnets to focus the beam and correction magnets (\Fref{Components-Circular-Accelerator}). Pulsed magnets are required for injection and extraction. Power supplies provide the magnet current. Radio-frequency (RF) cavities accelerate the beam; the RF power is provided by the RF system. Other systems include beam instrumentation and the control system. The vacuum system ensures very low pressure for the beam circulating in the vacuum chamber.

The beams are injected at low energy and the energy is increased while ramping the magnetic field. The particles are accelerated by RF fields in RF cavities.

\begin{figure}
\centering
\includegraphics[width=0.6\linewidth]{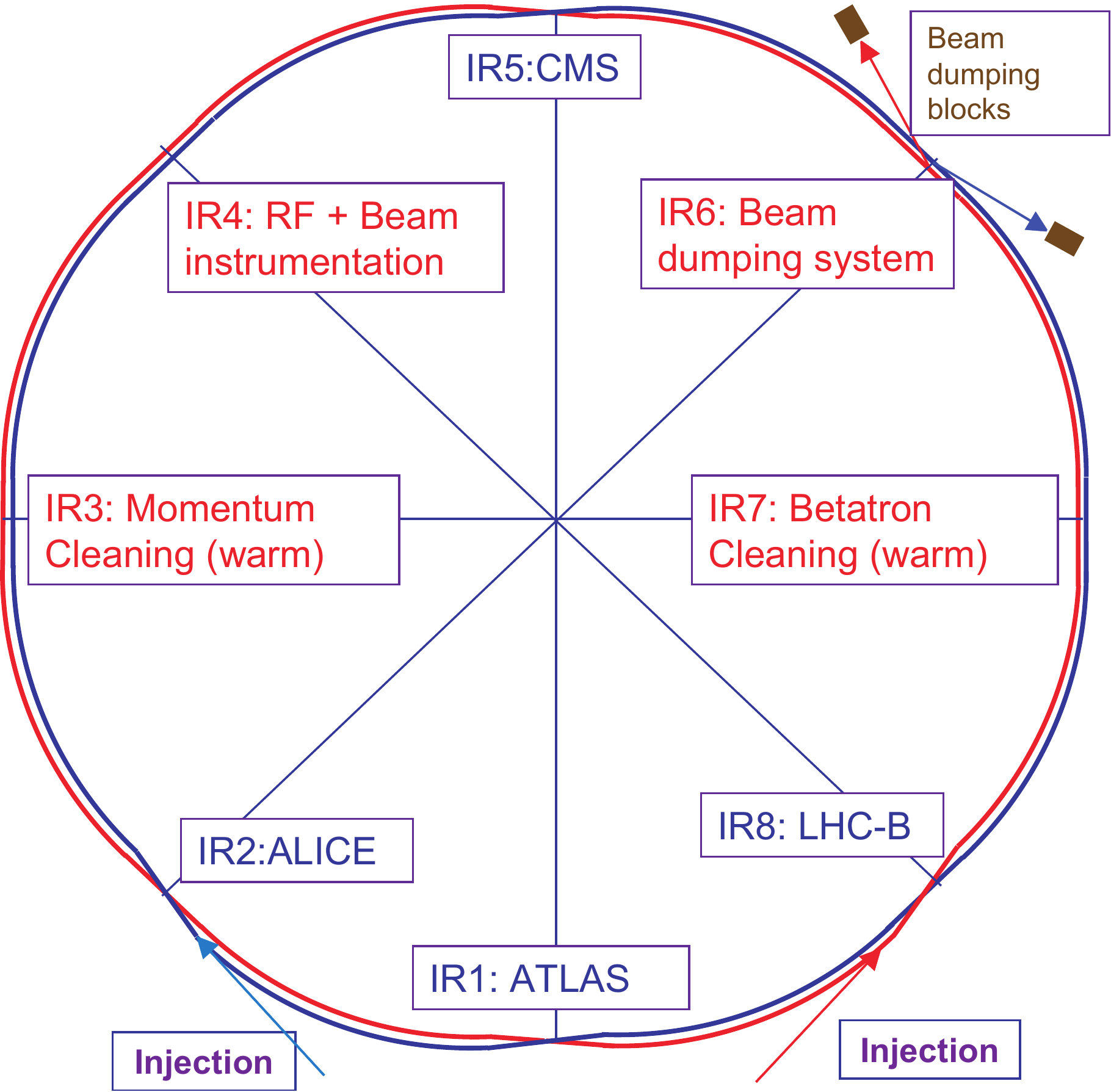}
\caption{Layout of LHC. Injection of the beams is via two 3~km long transfer lines from the SPS}
\label{LHCLayoutx}
\end{figure}

The layout of the LHC is shown in \Fref{LHCLayoutx} \cite{TheLHCStudyGroup1995}. It has eight arc sections and eight straight sections (or insertions) for experiments and accelerator systems. In four straight sections physics experiments are installed; in IR2 and IR8 these are together with the systems for injecting the beams coming from the SPS via two transfer lines. Protection systems are an essential part of the layout. Three insertions are used for elements related to machine protection: one insertion for the beam dumping system and two insertions for the collimation systems. Around the circumference more than 3600 beam loss monitors are installed. If one monitor detects beam losses above a predefined threshold, a signal is transmitted via an interlock system to the extraction kicker magnets in the beam dumping system and the beams are extracted towards the beam dump blocks.

\begin{figure}
\centering
\includegraphics[width=0.8\linewidth]{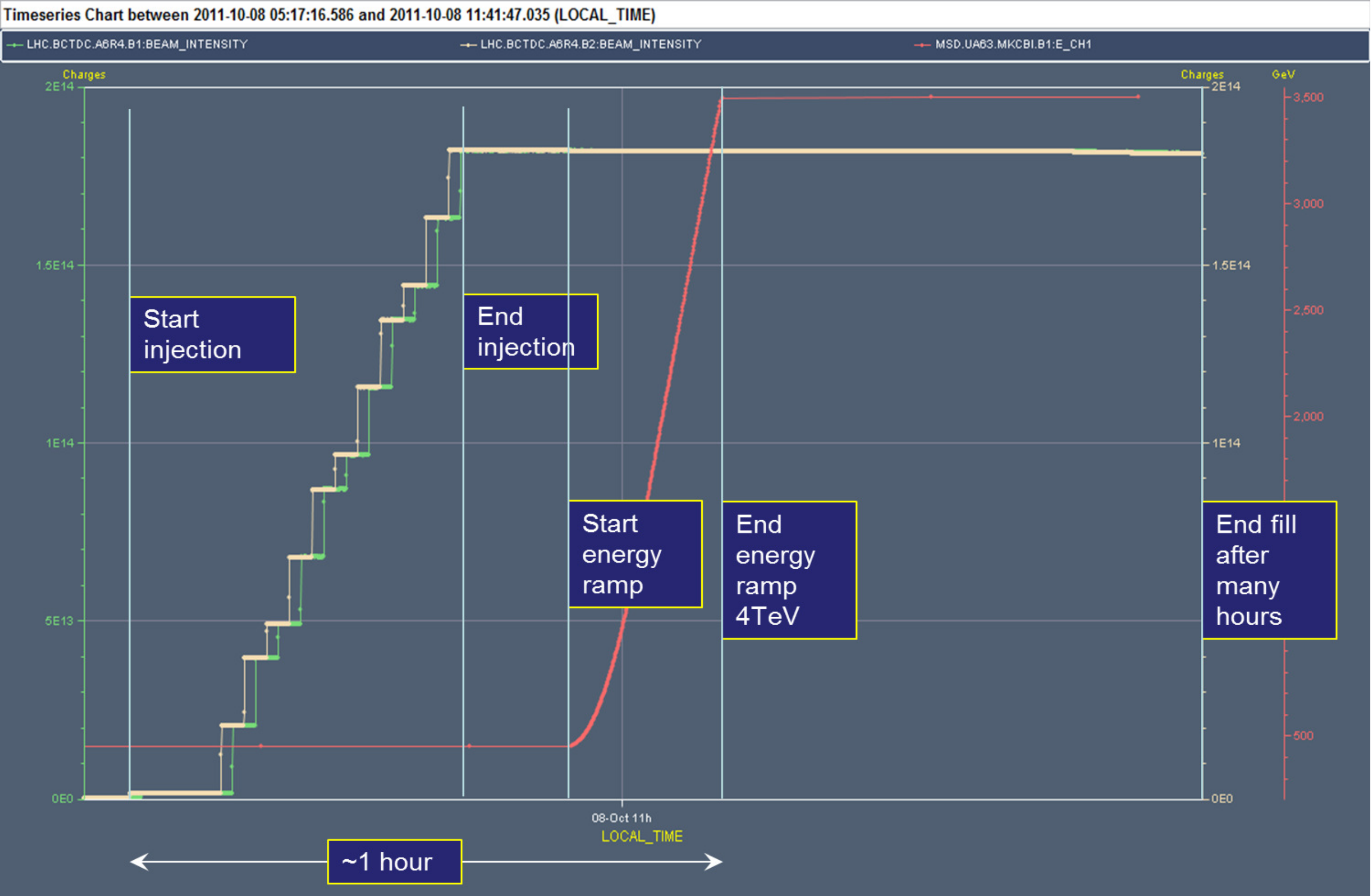}
\caption{Operational cycle of LHC, with injection at an energy of 450~GeV, energy ramp to 4~TeV (in 2012) and colliding beams.}
\label{LHC-operational-cycle-real-data}
\end{figure}

A typical operational cycle is shown in \Fref{LHC-operational-cycle-real-data}. Beams are injected in batches with up to 288 bunches from the SPS to LHC at an energy of 450~GeV. It takes some 10~min to fill the two beams. Then acceleration starts and the energy ramp takes about 20~min. At top energy, the beams are brought into collisions and collide for many hours for the data taking of the physics experiments (from a few hours to some tens of hours). At the end of the fill and in case of a failure the beams are extracted towards the beam dump blocks.

Three phases of beam operation related to machine protection are defined: injection, operation with stored beams and the extraction of the beams.

\section{Failures with an impact on the beam}

\subsection{Classification of failures}

In the first step different types of failures that can cause beam losses are identified and classified:

\begin{itemize}
\item hardware failure (trip of a power converter, magnet quench, AC distribution failure, object in vacuum chamber, vacuum leak, RF trip, kicker magnet misfire, etc);
\item control failure (wrong data, wrong magnet current function, trigger problem, timing system failure, feedback failure, etc);
\item operational failures (chromaticity/tune/orbit wrong values, etc);
\item beam instability (due to too high beam current/bunch current/e-clouds, etc).
\end{itemize}

The most important parameters for a failure are:

\begin{itemize}
\item time constant for beam loss after the occurrence of the failure;
\item probability of the failure occurring;
\item damage potential in case no mitigation is applied.
\end{itemize}

An accurate understanding of the time constant is required, since this determines the reaction time of the machine protection systems. The risk defined as ${\rm risk} = {\rm consequences} \times {\rm probability}$ is another important input determining the required reliability for the protection systems. For very high risk the protection systems must be extremely reliable.

\subsection{Time constant for failures}

The time constant for beam loss after a failure varies from nanoseconds to many seconds.

\textbf{Single-passage beam losses} in the accelerator complex have a time constant of a few nanoseconds to some tens of $\mu$s. In a circular accelerator such losses are related to failures of fast kicker magnets for injection and extraction. If other fast kicker magnets are present, for example for diagnostics, failures of such devices must also be considered. For failures of fast kicker magnets it is not possible to extract the beam or to stop the beam at the source, the particles will travel determined by the electromagnetic field along their path.

Single-passage beam losses are also an issue for any accelerator operating with pulsed beam. In between two pulses, equipment parameters can change (e.g. a magnet power supply can trip). During the following beam pulse, the beam would be mis-steered and can cause damage. This is typically the case for failures in a transfer line between accelerators (e.g. from SPS to LHC) or from an accelerator to a target station (target for secondary particle production or beam dump block). This is also an issue for linear accelerators operating with pulsed beams.

\textbf{Very fast beam losses} with a time constant in the order of 1 ms, e.g. multiturn beam losses in circular accelerators. Such losses can appear due to a large number of possible failures, mostly in the magnet powering system, with a typical time constant of about 1 ms to many seconds.

\textbf{Fast beam losses} with a time constant of 10~ms to seconds, due to many different effects. Beam instabilities in LHC are in general in this time range.

\textbf{Slow beam losses} take many seconds, e.g. due to non-optimized parameters, but also due to a failure.

Details for beam losses in circular and linear accelerators are presented in \cite{Kain2014,Kain2014a,Plum2014}.

\section{Particle distribution and aperture}

The evaluation of the consequences of a failure requires the analysis of the trajectories of the particles due to the failure, in particular the location where they will touch the aperture.

\begin{figure}
\centering
\includegraphics[width=0.8\linewidth]{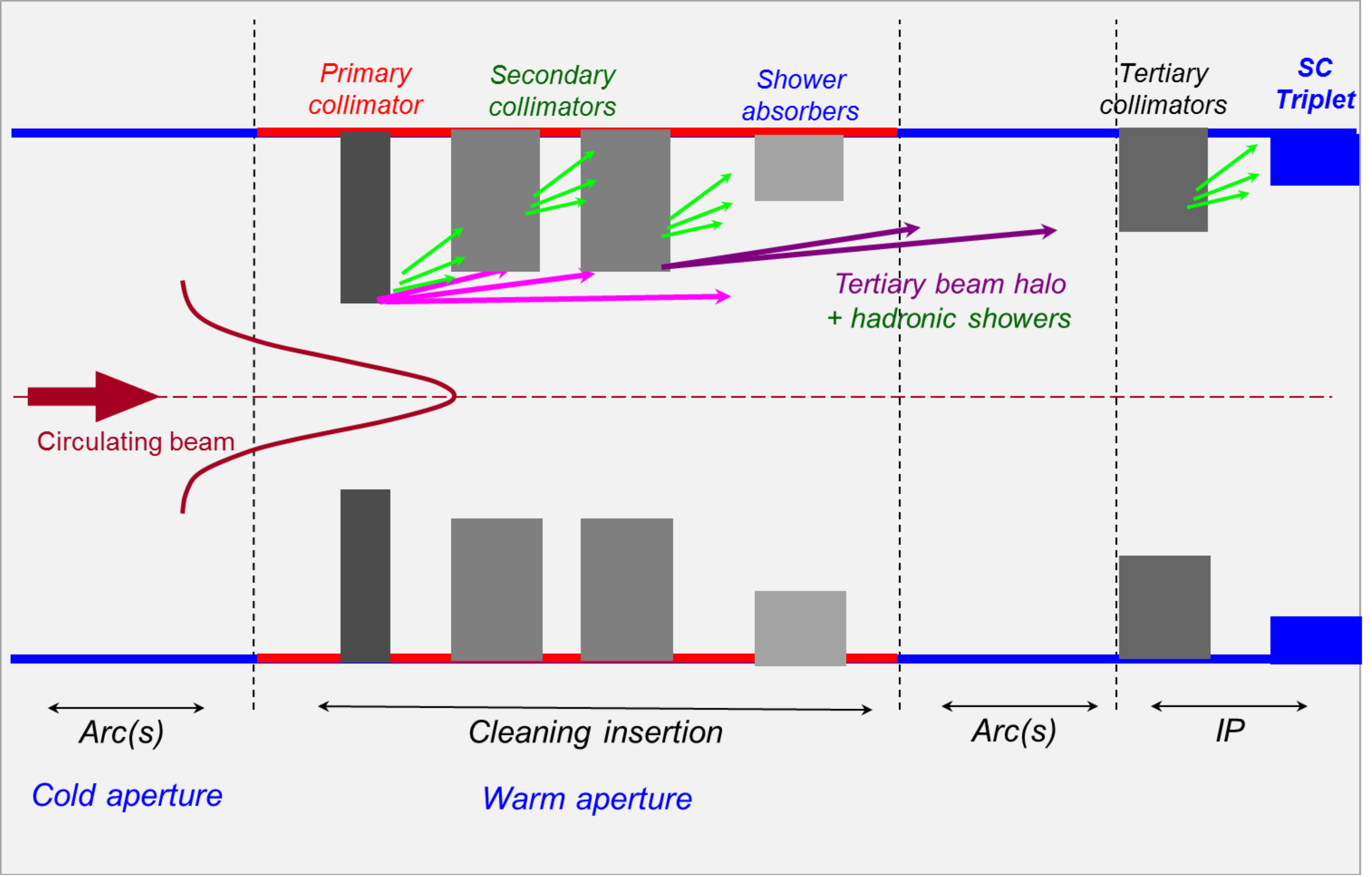}
\caption{Illustration of the layout of the beam cleaning system in LHC (see \cite{Redaelli2014})}
\label{Collimation-layout}
\end{figure}

In LHC, collimators are installed in two long straight cleaning insertions; they are always positioned to limit the aperture \cite{Redaelli2014}. Primary collimators are set to a position closest to the beams. After a failure, the emittance and therefore the beam size might increase, the closed orbit might change or both happen at the same time. If the amplitude of the betatron oscillation of a particle increases, the particle will first hit a collimator and not the superconducting magnets or other parts of the accelerator. In general, particles are expected to hit a primary collimator. Scattered particles and showers from the collision of the protons with material are absorbed in secondary or tertiary collimators (see \Fref{Collimation-layout}). A typical position of the primary collimator with respect to the beam centre is in the order of 6 $\sigma$, with $\sigma$ defined as the root mean square beam size. Details of the collimation system are presented in \cite{Redaelli2014}.

\begin{figure}
\centering
\includegraphics[width=0.8\linewidth]{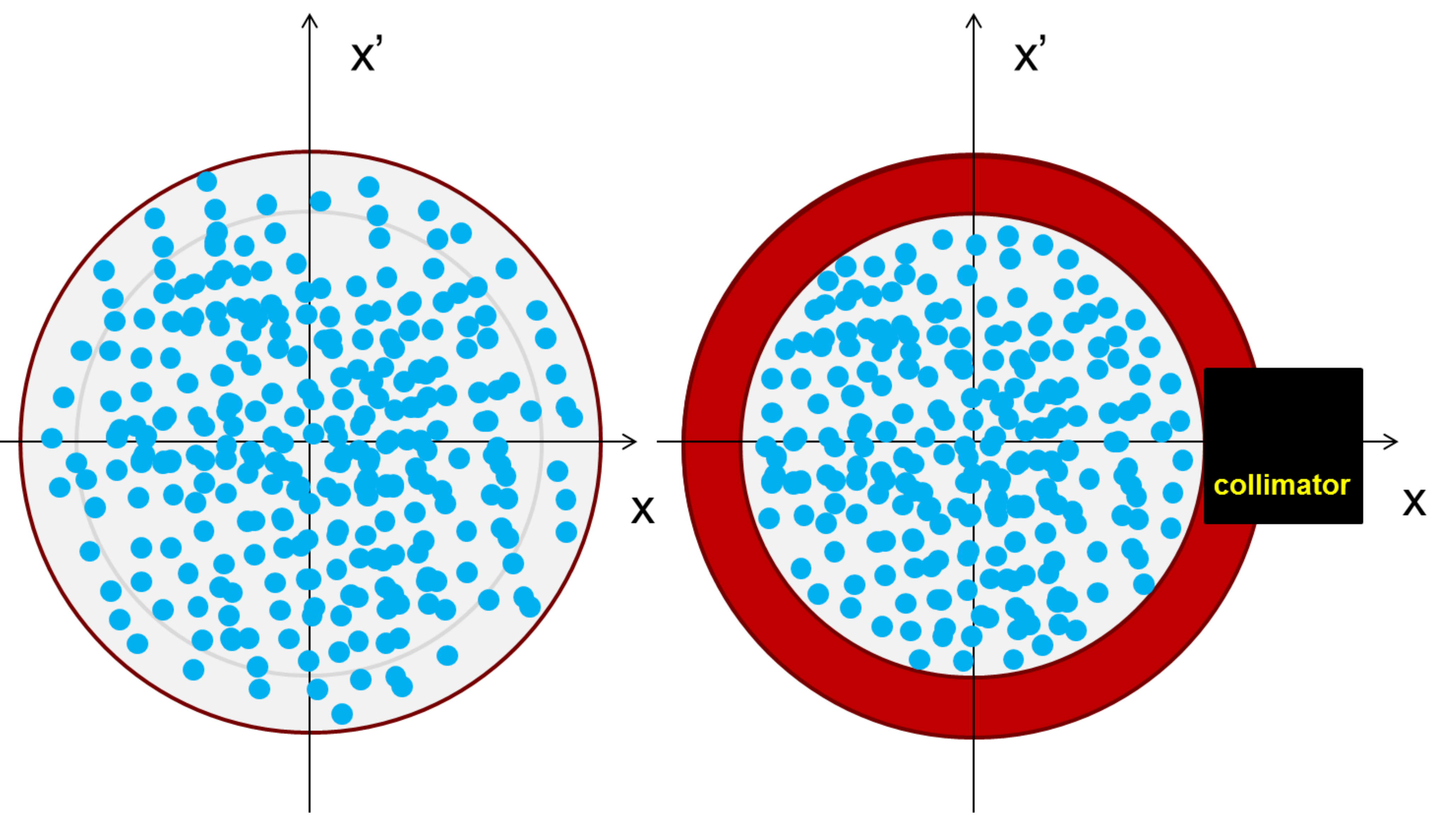}
\caption{Illustration of the phase-space reduction by a collimator. On the left, the particle distribution in phase space is shown. On the right, a collimator is moved into the beam and all particles in the red circle are scraped.}
\label{Phase-Space-Reduction}
\end{figure}

If a collimator is moved closer to the beam, all particles with amplitudes above the value defined by the collimator jaw position will be scraped away (\Fref{Phase-Space-Reduction}). When the entire beam starts to move, e.g. in case of a magnet trip, the same happens and all particles with amplitudes larger than the position of the collimator will hit the collimator within a few turns.

\begin{figure}[tb]
    \centering
    \includegraphics*[width=0.75\linewidth]{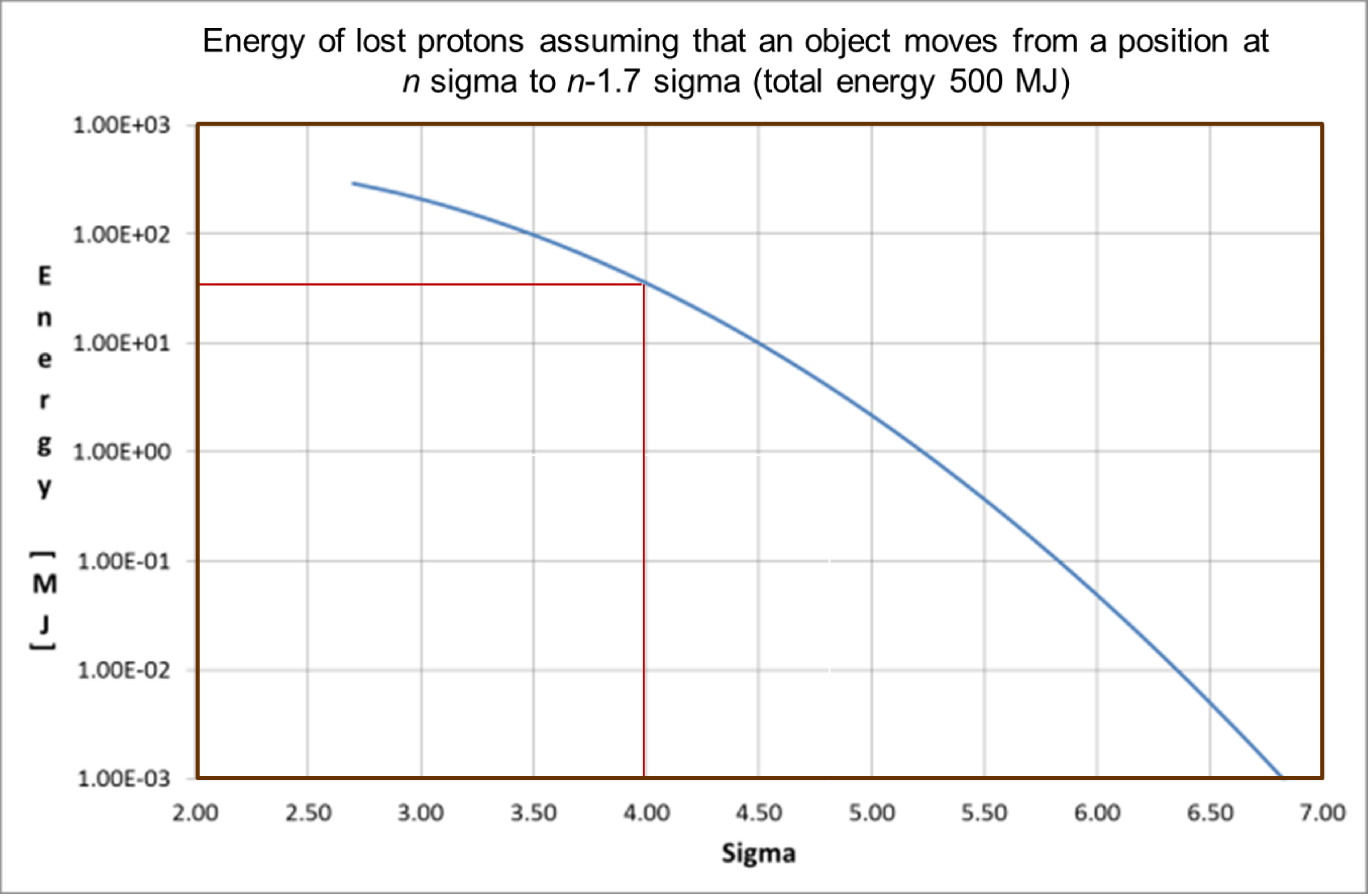}
    \caption{Energy loss assuming that an object cuts into the beam tail by 1.7~$\sigma$}
    \label{Energy-distribution-crab-cavity-failure}
\end{figure}

A Gaussian particle distribution is assumed together with a collimator at a position corresponding to 4$~\sigma$. In case of a failure and a fast displacement of the beam by, say, 1.7~$\sigma$, all particles above an amplitude of 2.3~$\sigma$ would hit the collimator jaw (see \Fref{Phase-Space-Reduction}). If the energy stored in the beam corresponds to, say, 500~MJ, the energy loss would correspond to 35~MJ and the collimator would explode. For a collimator at 5~$\sigma$ and the same fast displacement the energy loss is 2.2~MJ and for a collimator at 6~$\sigma$ the energy loss is less than 0.1~MJ. The energy loss as a function of collimator setting in case of such failure is shown in \Fref{Energy-distribution-crab-cavity-failure}.

\begin{figure}
\centering
\includegraphics[width=0.8\linewidth]{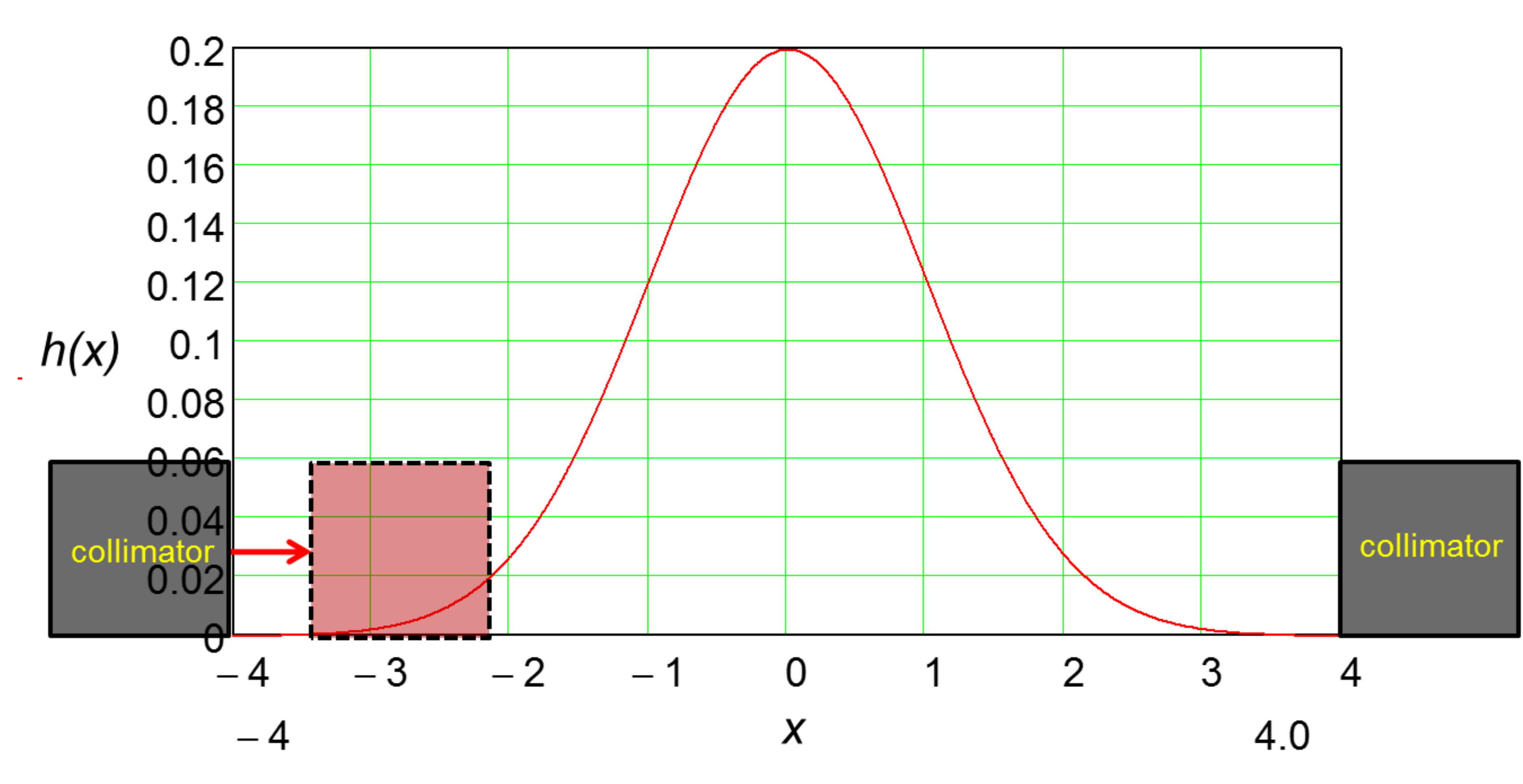}
\caption{Gaussian beam distribution with a collimator position at 4 $\sigma$, moving by 1.7 $\sigma$ to 2.3 $\sigma$. At 4 $\sigma$, about 99\% of the particles survive; at 2.3 $\sigma$, this value is about 93\%.}
\label{Collimation-moving-in}
\end{figure}

\section{Mechanisms for creating beam losses: overview}

\noindent \textbf{Magnetic fields}: In a circular accelerator magnetic fields are the dominant elements that determine the particle trajectories, either normal conducting  or superconducting. Dipoles magnets provide the deflection of the particles, quadrupoles are installed for focusing, sextupoles for compensating the chromaticity and higher order multipole magnets for various reasons.

\noindent \textbf{Electric fields}: All circular accelerators use RF cavities with longitudinal electrical fields for acceleration and bunching of the beams. In some accelerators electrical fields are applied for transverse deflection. Examples are electrostatic separators to separate beams that have the same charge, e.g. at LEP and the SPS proton--antiproton collider. Transverse feedback systems are also frequently using devices that produce an electrical field. A new method for tilting the beam to deflect particles in a bunch as a function of the particle position within the bunch is using RF cavities (so-called crab cavities), e.g. for the B-factory at KEK and proposed for HL-LHC \cite{Yee-Rendon2014}.

\noindent \textbf{Beam instabilities}: The bunch charge and the related electromagnetic fields can lead to beam instabilities. There are different types of instabilities acting on the particles within the bunch and on following bunches that can cause beam losses. At LHC the typical time constant for the growth of beam instabilities is in the order of several tens of ms to many seconds.

\noindent \textbf{Obstructions in the beam pipe}: There can be a piece of matter inside the beam pipe or the gas pressure can be much higher than the nominal value. The residual gas pressure in the vacuum pipe is in the order of $10^{-6}$ to $10^{-12}$ mbar. For beams that are circulating in the accelerator for many hours, the pressure should be less than, say, $10^{-8}$ mbar. In case of a vacuum leak or other effects (such as electron clouds) increasing the pressure, particles would collide with the gas preventing efficient operation. Beam losses risk quenching superconducting magnets and  activating accelerator equipment. In most accelerators equipment is installed that can move into the beam pipe, such as vacuum valves and screens for the observation of the beam profile. If such elements are accidentally moved into a high-intensity beam, the beam will damage the equipment and hadron showers from the interaction of the beam with the obstructing material can cause further collateral damage.

\section{Magnetic and electrical fields}

When a dipole magnetic field in the accelerator slowly changes and deviates from the nominal field, the closed orbit changes. A change is considered to be slow when the orbit changes by 1 $\sigma$ in several turns. Fast-changing dipole fields (kicker magnets) introduce betatron oscillations. Quadrupole fields change the betatron tune and the optics and therefore the beam size around the accelerator. This might drive the beam onto resonances and cause beam instabilities. Sextupole fields change the chromaticity and might also drive the beam onto resonances and cause instabilities.

For LHC, a Ph.D. study \cite{Gomez-Alonso2009} showed that a failure of normal-conducting dipole magnets close to the experiments in IR1 and IR5 has the fastest impact on the beam (excluding kicker magnets). The failure cases were defined and the trajectories of the particles after the failure were calculated. This allowed the calculation of the time between the onset of the failure and the particles touching the aperture.

\subsection{Very fast beam losses: failures in normal-conducting magnet circuits}

We assume that the magnetic field is proportional to the magnet current, a slightly pessimistic assumption since eddy current in the vacuum chamber will slow down fast changes of the magnetic field. The relevant magnet parameters are inductance $L$ and resistance $R$; the power converter parameters are the current $I(t)$ and voltage $V(t)$. Magnets rarely fail; a magnetic field error is in general caused by a magnet current error. There are many failure modes for such error, due to a failure in the electrical supply, the power converter itself, after a quench, water cooling problems, controls or operation.

A power converter is a very complex device, but modelling a power converter failure for the purpose of machine protection considerations can be simplified. We assume nominal operation at constant current with the nominal voltage $V_{\rm nom}$ and the resistance $R$. The current is given by Ohm's law:

\begin{equation}
I(t) = V_{\rm nom} / R.
\end{equation}

\noindent We consider some failure scenarios.
\begin{itemize}
\item The power supply trips and voltage goes to zero (this is the most likely failure, e.g. during a thunderstorm or due to other reasons).
\item The control system requests the power supply to provide maximum voltage, possibly with opposite polarity $V_{\rm fail}$. The effect on the current as a function of time can be larger than for a trip of the power converter.
\end{itemize}

With $\tau=L/R$, the current after the failure is given by

\begin{equation}
I(t) = I_{\rm nom} \cdot \left({\rm e}^{-t/\tau} + \frac{V_{\rm fail}}{V_{\rm nom}} \cdot (1-{\rm e}^{-t/\tau})\right).
\end{equation}

A magnet with the field $B$ and a length $L$ deflects a particle with the energy $E$ by an angle

\begin{equation}
   \alpha = \frac{B \cdot L}{E} \cdot c \cdot e_0.
\end{equation}

The change of the closed orbit as a function of the deflection angle $\alpha$ in the horizontal plane is given by

\begin{equation}
   x = \frac{\sqrt{\beta_1 \cdot \beta_2}}{2 \cdot \sin(\pi \cdot Q)} \cdot \alpha,
\end{equation}
with $\beta_1$ and $\beta_2$ the values of the betatron function at the location of the magnet and the location of the observation point, respectively, and $Q$ the betatron tune.

\begin{figure}[tb]
    \centering
    \includegraphics*[width=0.9\linewidth]{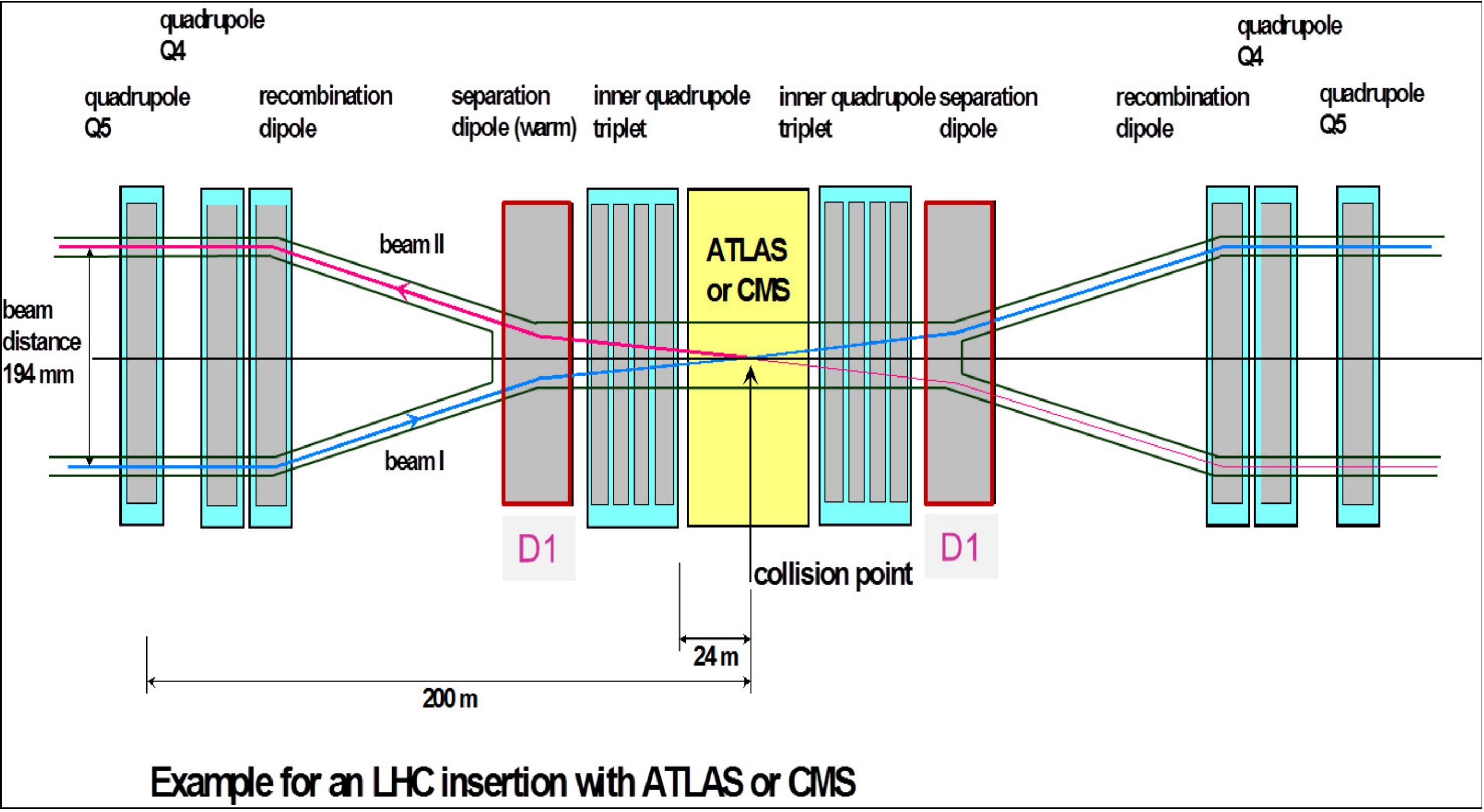}
    \caption{Illustration of a LHC insertion with the normal-conducting D1 magnets to separate the beam left and right from the collision points.}
    \label{LHC-insertion-with-D1}
\end{figure}

Resistive magnets have a high resistance and a low inductance compared to superconducting magnets and their current decay can be fast. If the magnet is installed at a position with high $\beta$ function, the orbit change due to the failure is fast. Such magnets are installed in two of the LHC insertions to separate the beams (so-called D1 magnets, see \Fref{LHC-insertion-with-D1}). The LHC beams are brought together to collide in a common region. Over ~260 m the beams circulate in one vacuum chamber with parasitic encounters (when the spacing between bunches is small enough). The D1 magnets separate the two beams.

The inductance for 12 magnets powered in series is 1.7~H and the resistance 0.78 $\Omega$. The nominal field at 7~TeV is 1.38~T; this yields a deflection angle of $\alpha = 2.41$ mrad. With a time constant in case of a power converter trip of $\tau=2.53$ s, the magnetic field error after 10 turns (this corresponds to 89 $\mu$s) is $\Delta B/B = 3.52\times 10^{-4}$. The deflection angle is $\alpha = 0.848$ $\mu$rad. The $\beta$ function at the D1 magnets is 4000~m. The change of the beam position at a location with a $\beta$ function of 100~m is 0.32~mm after a period of 10 turns. This change of the orbit corresponds to a change of 1.4 $\sigma$, when assuming nominal emittance.

\begin{figure}[tb]
    \centering
    \includegraphics*[width=0.9\linewidth]{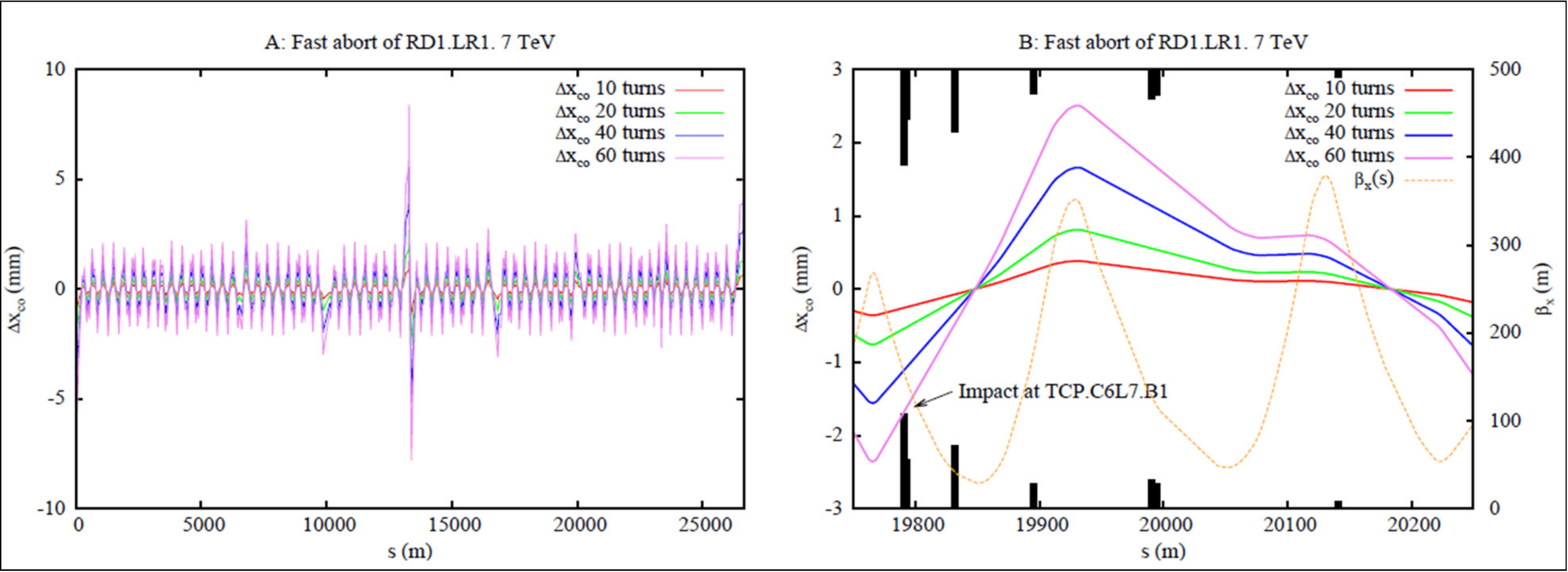}
    \caption{Change of closed orbit after a trip of the power converter for the D1 magnets. The left-hand graph shows the closed orbit around LHC and the right-hand graph the orbit in the collimation section \cite{Gomez-Alonso2009}.}
    \label{LHC-orbit-for-D1-failure}
\end{figure}

This type of failure as well as many other powering failures was simulated with the beam optics and particle tracking program MADX and the result is shown in \Fref{LHC-orbit-for-D1-failure}.

Apart from failures of the extraction and injection kickers, the impact of a failure of the D1 magnet was identified as the fastest mechanism for creating beam losses. The orbit starts to move rapidly by 1 $\sigma$ in about 0.7 ms. In 10 ms the beam would move by 14 $\sigma$, already outside of the aperture defined by the collimators. If such failure happens, the beam has to be extracted in a very short time. The probability for such failure during the lifetime of LHC is high. The consequences without protection would also be catastrophic, destroying the entire collimation system and possibly causing further damage. The protection needs therefore to be very fast and reliable. Basic parameters of the LHC MPS were designed to cope with this type of failure, in particular the reaction time of the MPS. The failed beam must be detected and the beam must be extracted in less than 1~ms. The detection of the failure is performed by several different systems (diverse redundancy).

\begin{itemize}
\item Detection of the failure of a wrong magnet current is challenging, since a fast detection of a current change on the level of $10^{-4}$ is required. This is done with specifically designed electronics (FMCM = fast magnet current monitor) \cite{Werner2005}.
\item Beam loss monitors detect losses when the beam touches the aperture and particles create a hadron shower (e.g. close to collimator jaws, but also other components) \cite{Dehning2014}.
\item In the future a fast beam current monitor will be used to measure the circulating current. In case of fast beam current changes exceeding a predefined threshold the beam will be extracted.
\end{itemize}

\subsection{Fast beam losses: failures in superconducting magnet circuits}

The parameters for superconducting magnets are very different from those of normal-conducting magnets. The inductance is high and the resistance low (determined only by the resistance of the normal-conducting cables between power converter and cryostat). In case of powering failure, the decay of the magnet field takes a long time (up to many hours). In case of a quench, the current decay depends on the mechanism that causes the quench. In general, the magnet starts to quench at a specific location and the quench spreads out. The resistance increases with time. Superconducting magnets require a magnet protection system that detects any quench, switches off the power converter and activates quench heaters. This signal is also provided to the interlock system that requests a beam dump.

\begin{figure}[tb]
    \centering
    \includegraphics*[width=0.9\linewidth]{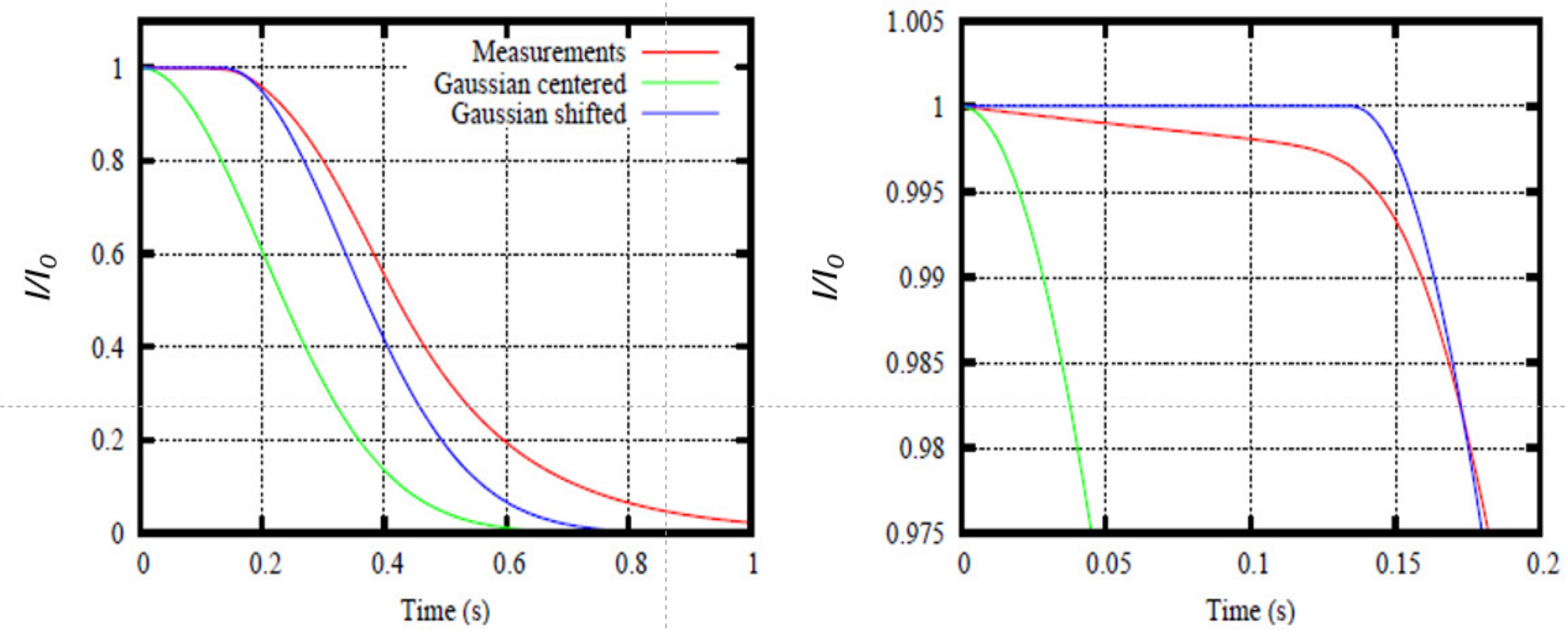}
    \caption{Current decay for a quench in one of the LHC superconducting dipole magnets. The measured current shows a small linear drop until the entire magnet is quenched and then follows a Gaussian decay. The characteristics of the linear drop depend on the evolution of the quench in each particular case and on the reaction time of the magnet protection system. In this particular case, the analytical approach with a time constant of 200~ms yields a decay that is slightly faster than the measured one \cite{Gomez-Alonso2009}.}
    \label{Current-dipole-quench}
\end{figure}

There is no analytical equation for the current after a quench as a function of time. In \Fref{Current-dipole-quench}, measurements from LHC magnets are shown. The current decay can be approximated by a Gaussian. In general, a quench is much less critical than the trip of the normal-conducting D1 magnet.

\subsection{Failures of the transverse damper}

The transverse damper is used to damp injection oscillations and instabilities. It also has several other applications: cleaning of the particle free abort and injection gaps, blowing up the beam for loss maps and aperture studies and generating beam losses for quench tests. In the future it will be used as a diagnostic tool to record bunch by bunch oscillations and for betatron tune measurement.

A worst-case calculation of the oscillation amplitude that the damper can  produce follows. The damper creates an electrical field for deflecting the particles and can deflect the beam with an angle of $\alpha =2$ $\mu$rad per passage at 450~GeV. In the worst case, the deflection of the transverse damper can add up coherently.

We assume that the betatron functions at the damper and at an observation point are $\beta=100$ m. The particles are deflected by a single kick to an amplitude of $x_{\rm d450}=\alpha \cdot \beta$ that corresponds to 0.2~mm, and at 7~TeV to 0.013~mm.

We assume a normalized emittance of $\epsilon_n=3.75 \times 10^{-6}$ m. The beam size at a location with a $\beta$ function of 100~m yields $\sigma=0.24$ mm. In the worst case of coherent oscillation, an amplitude of 1 $\sigma$ is reached after 18 turns, slightly less critical than a failure of the D1 magnet.

\section{Failures of fast kicker magnets}

Fast kicker magnets are required for injection of beam into LHC and for extraction towards the beam dump block. Failures of kicker magnets cannot be mitigated by active protection systems. If the beam is mis-steered due to a kicker magnet failure, the only option is to install beam absorbers for capturing the beam. There are several failure modes for kicker magnets.

\begin{itemize}
\item The deflecting angle is wrong.
\item The time when the kicker fires is wrong (too early, too late, too short, too long).
\item The injection kicker deflects the beam when it is not intended to operate, e.g. when LHC is operating above injection energy.
\end{itemize}

Some failure modes can be avoided by interlocks and operational procedures. As an example, the injection kicker should never deflect the beam when the accelerator is not at injection energy; therefore, after starting the ramp the injection kicker is switched off. There are some failure modes that cannot be avoided, such as a flash-over of a kicker; such failures need to be mitigated in order not to damage equipment.

\begin{figure}[tb]
    \centering
    \includegraphics*[width=0.8\linewidth]{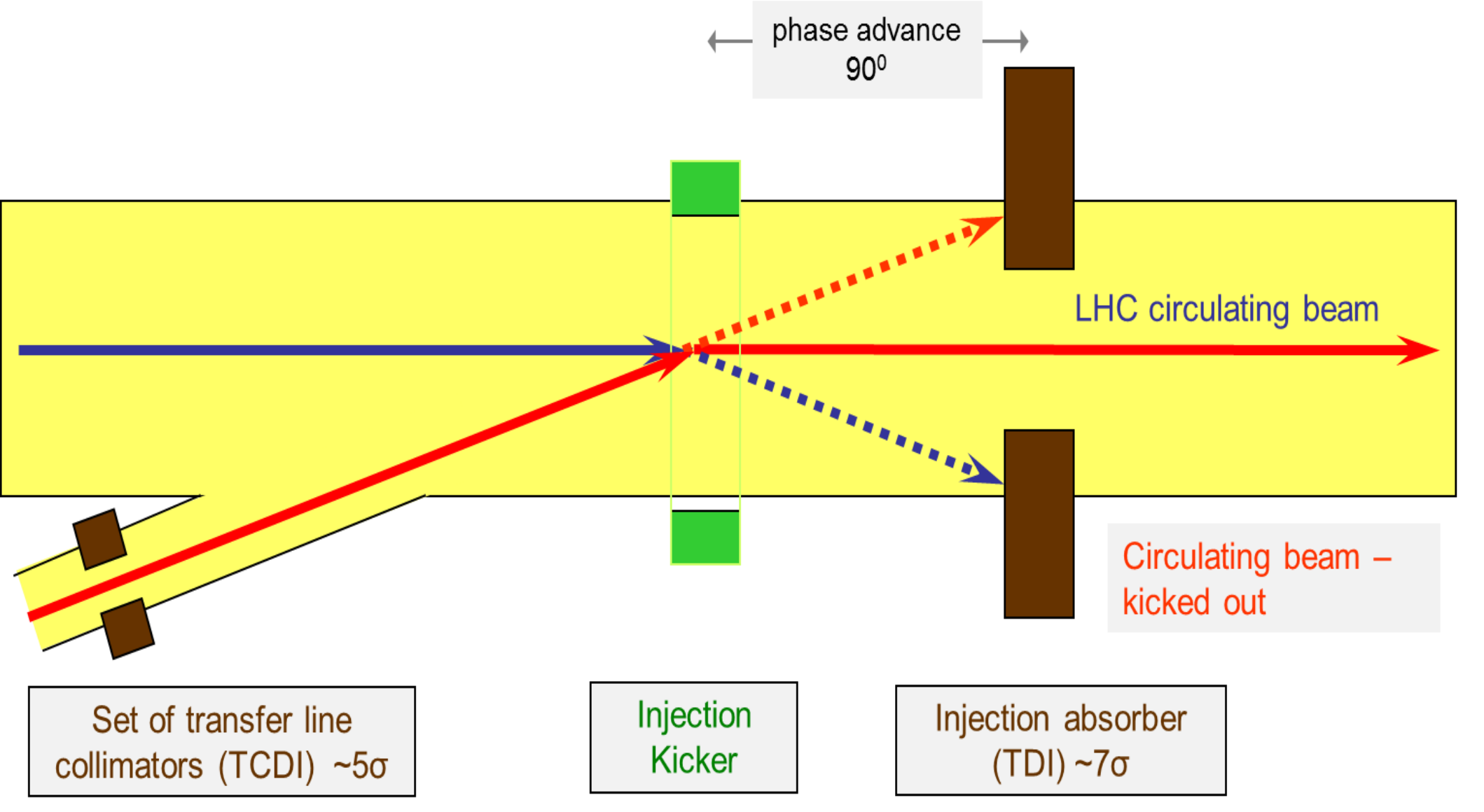}
    \caption{Illustration of failures at injection and related protection absorbers. One failure is the injection kicker not deflecting the beam that would travel straight into an absorber protecting the aperture. Another failure is the kicker deflecting the circulating beam that would travel to an absorber on the opposite side.}
    \label{Injection-Protection}
\end{figure}

At injection, a batch of 288 bunches is injected always with the same energy of 450~GeV \cite{Kain2014}. The injection elements and kicker and septum magnets must always have the same strength. The kicker magnets have a very short pulse, deflecting by a small angle. The septum magnet is a DC magnet that can be slowly pulsed, with no magnetic field acting on the circulating beam. The energy stored in a batch with 288 bunches injected into LHC is about 2~MJ; in case of accidental release the beam would seriously damage LHC equipment. Injection happens very frequently, in order to fill each of the two beams with up to 2808 bunches. An example for the elements of injection protection is shown in \Fref{Injection-Protection}. Two failure cases are considered.

\begin{itemize}
\item The beam is transferred via the transfer line, but the injection kicker for deflecting the beam onto the closed orbit fails. The beam travels further and would damage equipment. An absorber is installed at a location with a betatron phase advance of 90~degrees to absorb the energy.
\item The timing of the kicker pulse is wrong and the circulating beam is deflected. A second absorber is installed to absorb the energy of bunches that are deflected.
\end{itemize}

\begin{figure}
\centering
\includegraphics[width=0.8\linewidth]{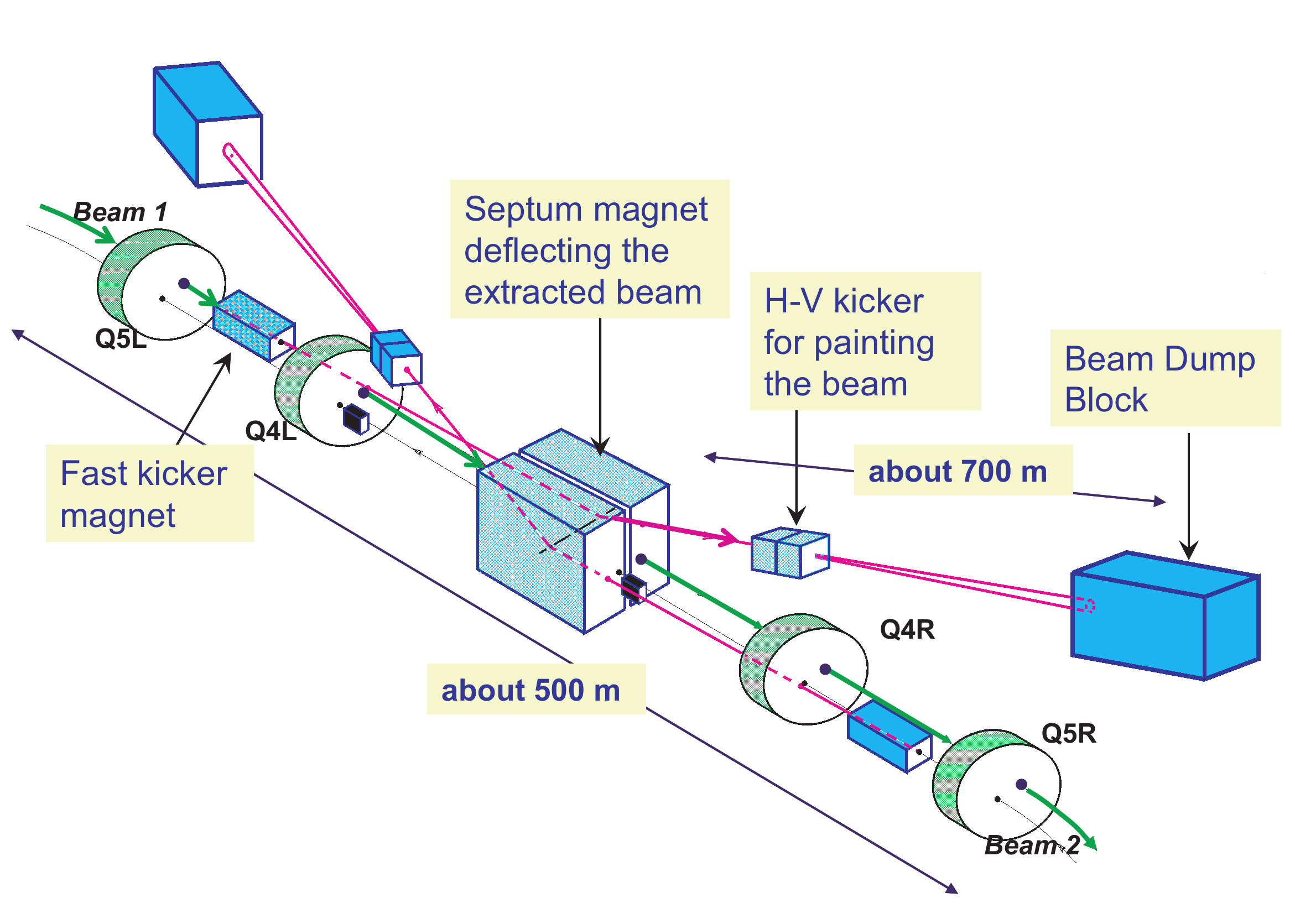}
\caption{Layout of the beam dumping system}
\label{BeamdumpLayout}
\end{figure}

One of the most critical components of the machine protection system is the beam dumping system (see \Fref{BeamdumpLayout}). One set of kicker magnets deflects the beam, septum magnets increase the deflection angle and a second set of kicker magnets dilutes the beam that travels through a 800~m long transfer line to the beam dump block.

Extraction can happen at any energy from 450~GeV to 7~TeV. The energy stored in the beam is up to 364 MJ. The strength of the kicker and septum field depends on the energy; the deflection angle must remain constant. The current of the main dipole magnets in four out of eight sectors of the LHC is measured and used to ensure the correct tracking of the kicker and septum magnet strengths. For a correct extraction, the kicker must fire synchronised with a particle free gap (during the kicker rise-time there are no particles), in order to deflect all bunches with the nominal angle. The duration of the kicker pulse must also be correct.

\section{Failure modes for magnets to mis-steer the beam}

There are many failure modes that can result in a wrong magnetic field:

\begin{itemize}
	\item failure of a power converter, water cooling or quench of a magnet;
	\item wrong command entered by an operator (e.g. request for angle change of 0.01 mrad instead of 0.001 mrad);
	\item timing event to start current ramp does not arrive or arrives at the wrong time;
	\item control system failure (data to power converter not sent or sent incorrectly);
	\item wrong conversion factor (e.g. from angle to power converter current);
	\item feedback system failure;
	\item failure of beam instrumentation, e.g. beam position monitor (BPM).
\end{itemize}

\begin{figure}
\centering
\includegraphics[width=0.8\linewidth]{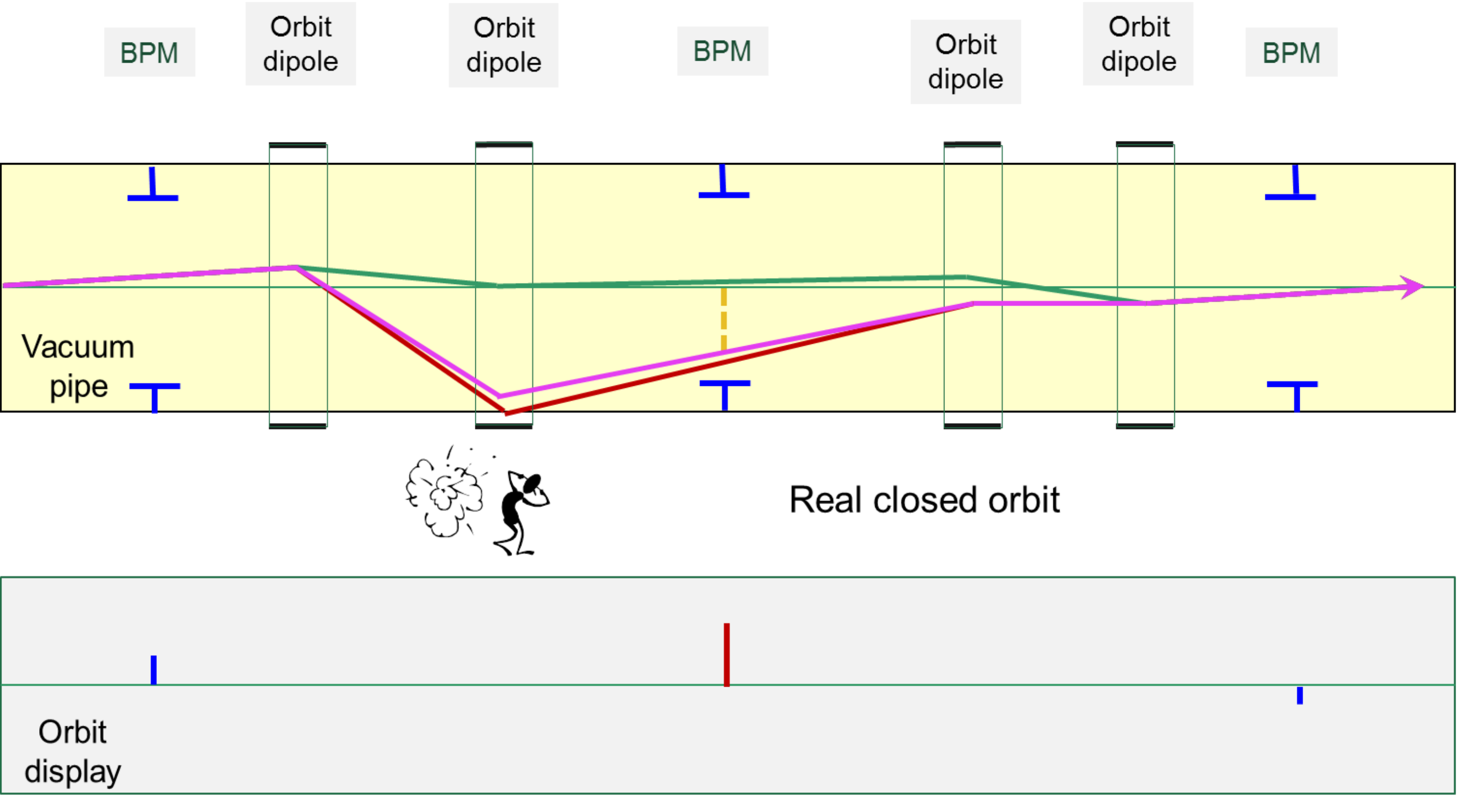}
\caption{The lower graph shows the reading of the orbit display, with one BPM always indicating the same wrong value. Initially, the orbit is correct. Then the BPM becomes faulty and the feedback system moves the beam by applying an orbit bump to correct for the offset until the beam touches the aperture.}
\label{Orbit-Problem-one-BPM-failure}
\end{figure}

One failure that is not obvious but must be considered is due to a wrong functioning of a beam position monitor. We assume that one of the beam position monitors is faulty and always provides the same wrong value, e.g. 1~mm, independent of the beam position (see \Fref{Orbit-Problem-one-BPM-failure}). In case an orbit feedback is operating, the feedback will try to correct the orbit at the position of the monitor by applying a closed orbit bump. The orbit deviation at this position will increase until the beam touches the vacuum chamber. Closed orbit bumps reduce the aperture and cannot be detected if a beam position monitor is faulty. An option to mitigate this fault is a software interlock reading the strengths of orbit corrector magnets.

\begin{figure}
\centering
\includegraphics[width=1.0\linewidth]{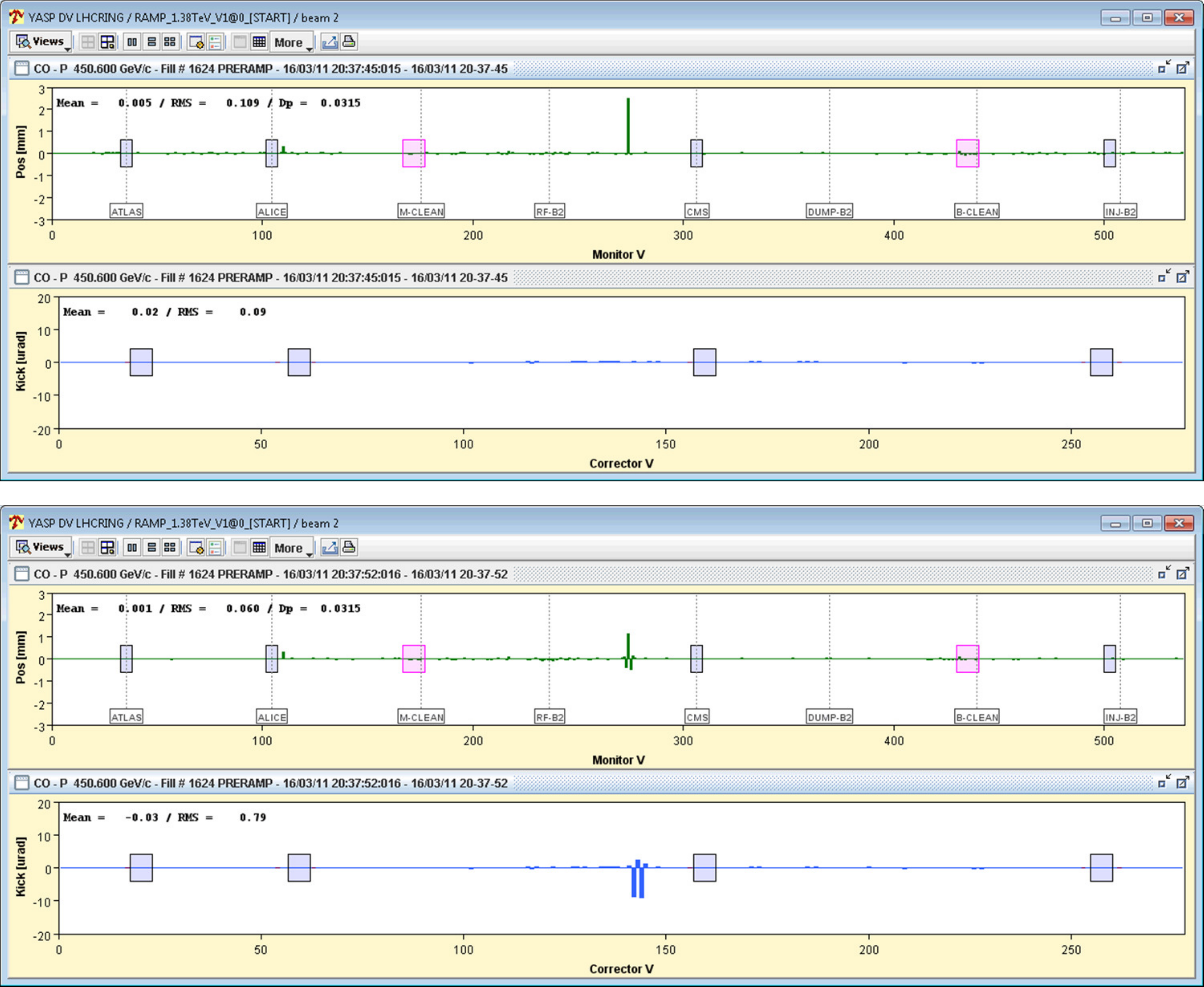}
\caption{Failure of a BPM at LHC. The upper graphs show the orbit before and after the failure; the lower graph shows the strengths of corrector magnets, with the bump building up. The data was taken with an interval of 7~s.}
\label{Orbit-Problem-one-BPM-data}
\end{figure}

Figure \ref{Orbit-Problem-one-BPM-data} shows that this really happens. The two graphs show orbit and strengths of corrector magnets at LHC for the vertical plane within an interval of 7~s; the data was taken in March 2011. In the first graph one BPM shows a large value and the feedback system started to correct the orbit with bumps. The bump was building up and finally the beam was lost.

\section{Objects that can block the beam passage}

Around an accelerator there are many objects that can block the beam passage. Some equipment is designed to move into the beam pipe (movable devices), other equipment should never be in the path of the beam, but this might accidentally happen:

\begin{itemize}
\item vacuum valves, possibly a large number to isolate different accelerator sections;
\item collimators and beam absorbers;
\item beam instrumentation: screens for observation of the beam profile, mirrors to observe synchrotron light, wire scanners to measure the beam profile;
\item experiments, e.g. so-called Roman pots, particle detectors to observe small angle scattered particles from the collision points that can move close to the beam.
\end{itemize}

\noindent Elements that should never be in the beam pipe:

\begin{itemize}
\item RF fingers for ensuring a continuous path for the image current across bellows. Such fingers can bend into the vacuum chamber;
\item other material, e.g. left-over pieces in the beam pipe from activities that require opening the beam vacuum;
\item elements that are getting into the pipe due to a failure (e.g. during cool down in a superconducting accelerator or during operation);
\item gas above nominal pressure.
\end{itemize}

\subsection{Wire scanners}

Wire scanners are made out of a thin carbon or tungsten wire that moves through the beam. The particle shower or the secondary emission is recorded to measure the beam profile. With too high energy density of the beam the wire risks melting; protection of the instrument is required. Superconducting  magnets in the vicinity risk quenching, leading to downtime of the accelerator. The energy density depends on the beam intensity and on the beam size. For hadron beams the size decreases with energy. Consequences of such failure are minor (instrument not available and risk of quenching a magnet). The probability that an operator sends the wire through a high-intensity beam is high. Since the risk is relatively small, a protection system with a low protection level is acceptable to prevent the wire passing through a high-intensity beam.

\subsection{Collimators and beam absorbers}

Collimators and beam absorbers are essential elements for machine protection (see \cite{Redaelli2014}). Collimators protect, but cannot absorb a wrongly deflected high-intensity beam and therefore also need protection, in particular at injection and extraction. They should be at the correct position with respect to the beam. Usually the closest collimator jaws are at a position of about 6 $\sigma$ from the beam centre. At LHC, the position of the collimators depends on the energy, since the beam size decreased during the energy ramp. Most collimators are therefore moved closer to the beam during the energy ramp.

A complex interlock system ensures that the collimators are correctly positioned. A timing event is used to start the collimator movement when the energy ramp starts. The position is verified by an independent method comparing the actual position of each collimator with the required position stored in a separate table for each energy. This can also be done when the optics changes and collimator position needs to be adjusted. Roman pots must always  be outside the aperture defined by the collimator.

\section{Consequences of beam loss}

\subsection{Beam losses and magnet quenches}

\begin{figure}
\centering
\includegraphics[width=0.7\linewidth]{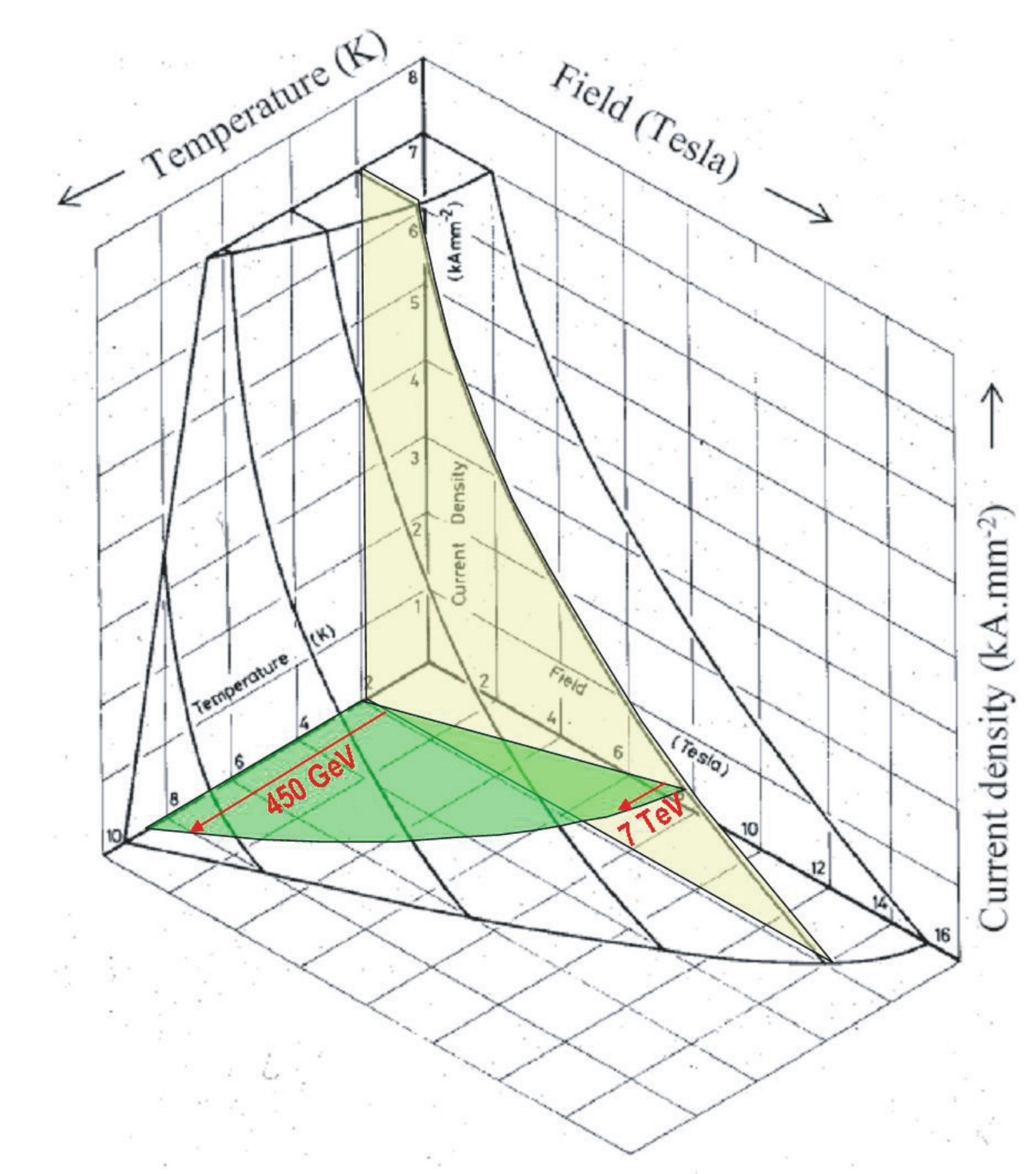}
\caption{Operational margin for NbTi magnets with magnetic field, current density and temperature. To operate with a magnetic field corresponding to an energy of 7~TeV, the operating temperature must be below 2~K.}
\label{SC_temp_margin}
\end{figure}

\begin{figure}
\centering
\includegraphics[width=0.6\linewidth]{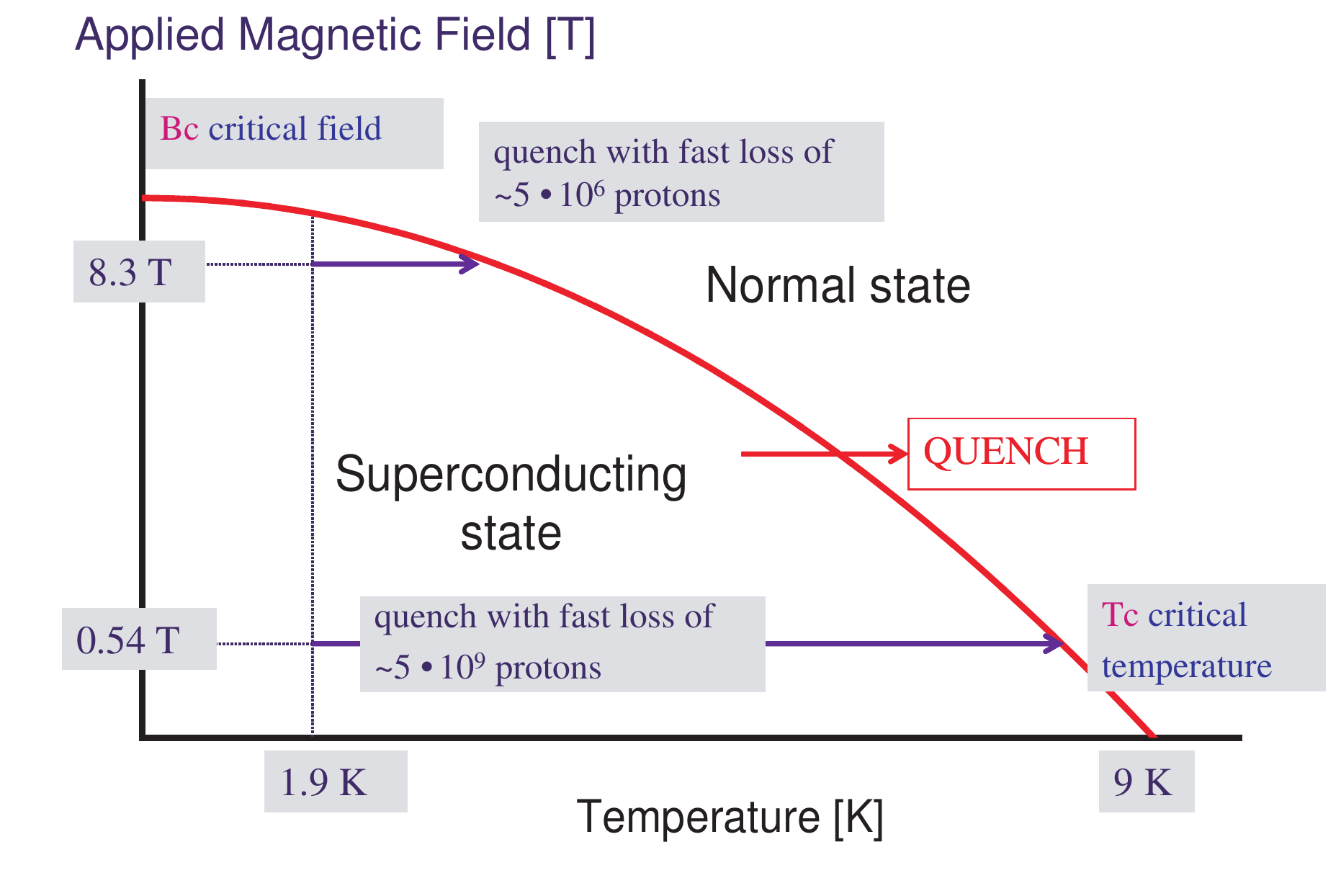}
\caption{Margin of a LHC superconducting magnet with respect to beam losses at injection and top energy. The absolute value of the energy deposition depends on the impact parameters; the numbers give the orders of magnitude.}
\label{OperationalMarginDipole}
\end{figure}

Superconducting magnets produce high field with high current density in the superconducting cables. The superconducting state of a magnet occurs only in a limited area of temperature, magnetic field and transport current density, depending on the superconductor (for LHC, NbTi conductors are used). Lowering the temperature enables better usage of the superconductor by broadening its working range. The operating parameters of LHC superconductors (NbTi) are shown in \Fref{SC_temp_margin}. Most of the LHC magnets operate at a temperature of 1.9~K.  When operating at low energy and therefore at low magnetic field, the temperature margin is relatively large and beam losses of some 100~J can be tolerated without quenching. For high energy, the margin is small and very small beam losses are sufficient to quench a magnet (see \Fref{OperationalMarginDipole}).

\subsection{Wrong deflection during extraction}

\begin{figure}[tb]
    \centering
    \includegraphics*[width=0.8\linewidth]{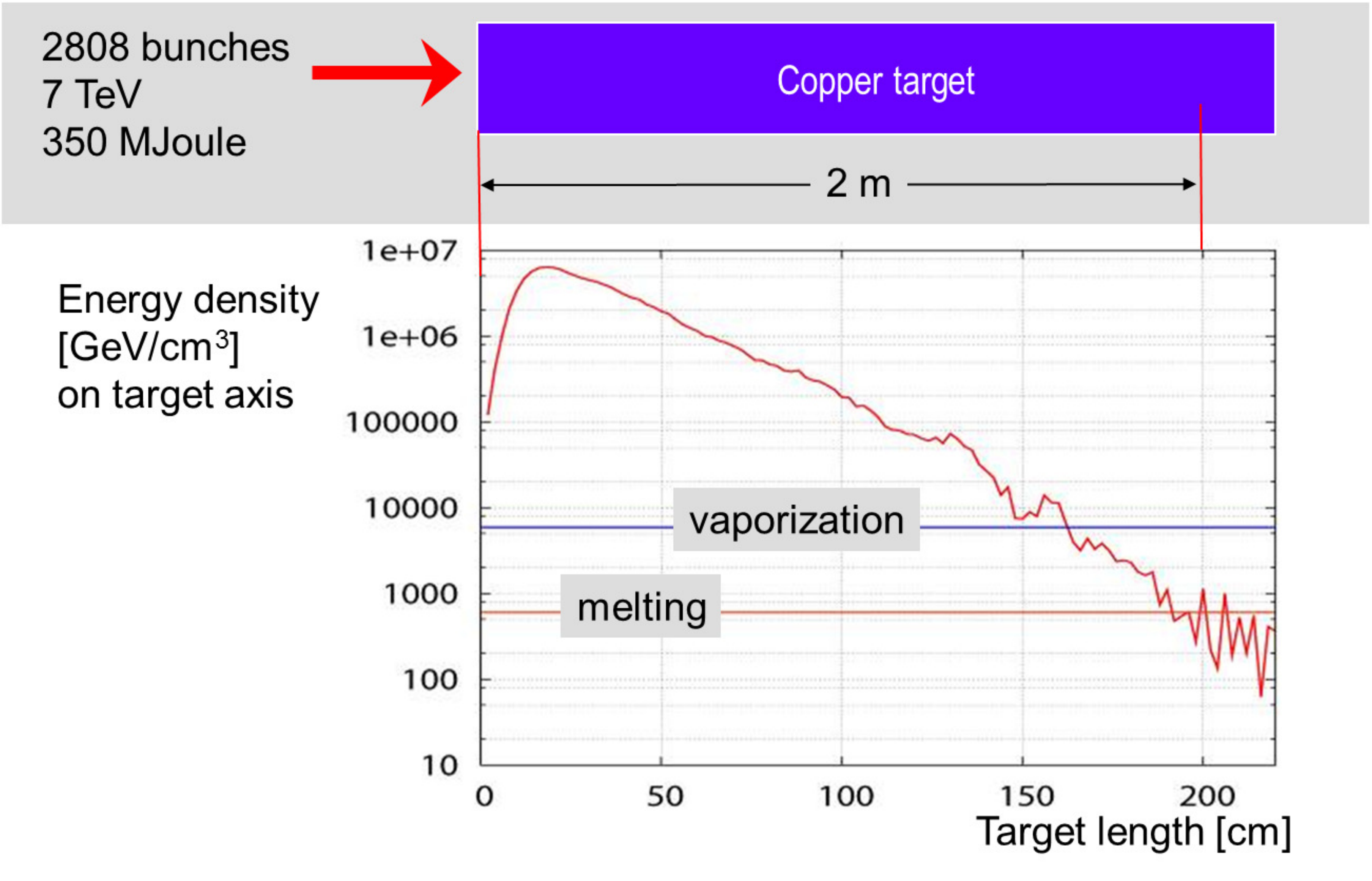}
    \caption{Energy density in a copper cylinder when a full 7 TeV LHC beam is deflected into a copper target}
    \label{Hydrodynamic-Tunneling}
\end{figure}

Several failures can lead to the beam deflected with non-nominal angle, e.g. if only one (out of 15) kicker magnets fires, or LHC operates at 7~TeV and the kicker magnets extract the beams with an angle corresponding to 450~GeV. Studies were performed to estimate the consequences of the full 7~TeV beam deflected into equipment such as a graphite collimator or a magnet \cite{Tahir2012}. A first indication is obtained by calculating the energy deposition in material with codes such as FLUKA \cite{fluka}. The result is shown for a copper target in \Fref{Hydrodynamic-Tunneling}. The energy density is far above the melting and vaporization points of the material.

The correct calculation needs to take the time structure of the beam into account. So-called hydrodynamic tunnelling of beam through the target becomes important after the impact of some 10 bunches. The first bunches arrive, deposit their energy and lead to a reduction of the target material density. Bunches arriving later travel further into the target since the material density is reduced. This effect has been already predicted for SSC \cite{Wilson-1993-PAC}. The calculation of hydrodynamic tunnelling is complex and performed in several steps. Typical parameters for the simulation are: 2808 bunches with $1.1\times 10^{11}$ protons, $\sigma=0.5$ mm and 25~ns bunch distance. The calculations predict a tunnelling depth of about 30~m for these parameters. Recently an experiment was performed to validate the simulation method \cite{Blanco-HB2012}.

\subsection{Damage levels for particle beams}

Commissioning of LHC starts with a low-intensity beam, in order to avoid any risk of damaging equipment. An obvious question is what beam parameters can lead to damage and what beam parameters are still safe. Relevant beam parameters are the particle type, the energy of the particles and the beam size. The answer to this question is far from being obvious. It depends not only on beam parameters, but also on the time distribution of the losses and the equipment that is exposed. In case of beam loss for more than a few ms cooling of the equipment has to be considered. Sensitive equipment such as particle detectors can be damaged by beam losses in the order of 1 J. Massive damage of equipment is not expected below some megajoules.

\begin{figure}[tb]
    \centering
    \includegraphics*[width=0.8\linewidth]{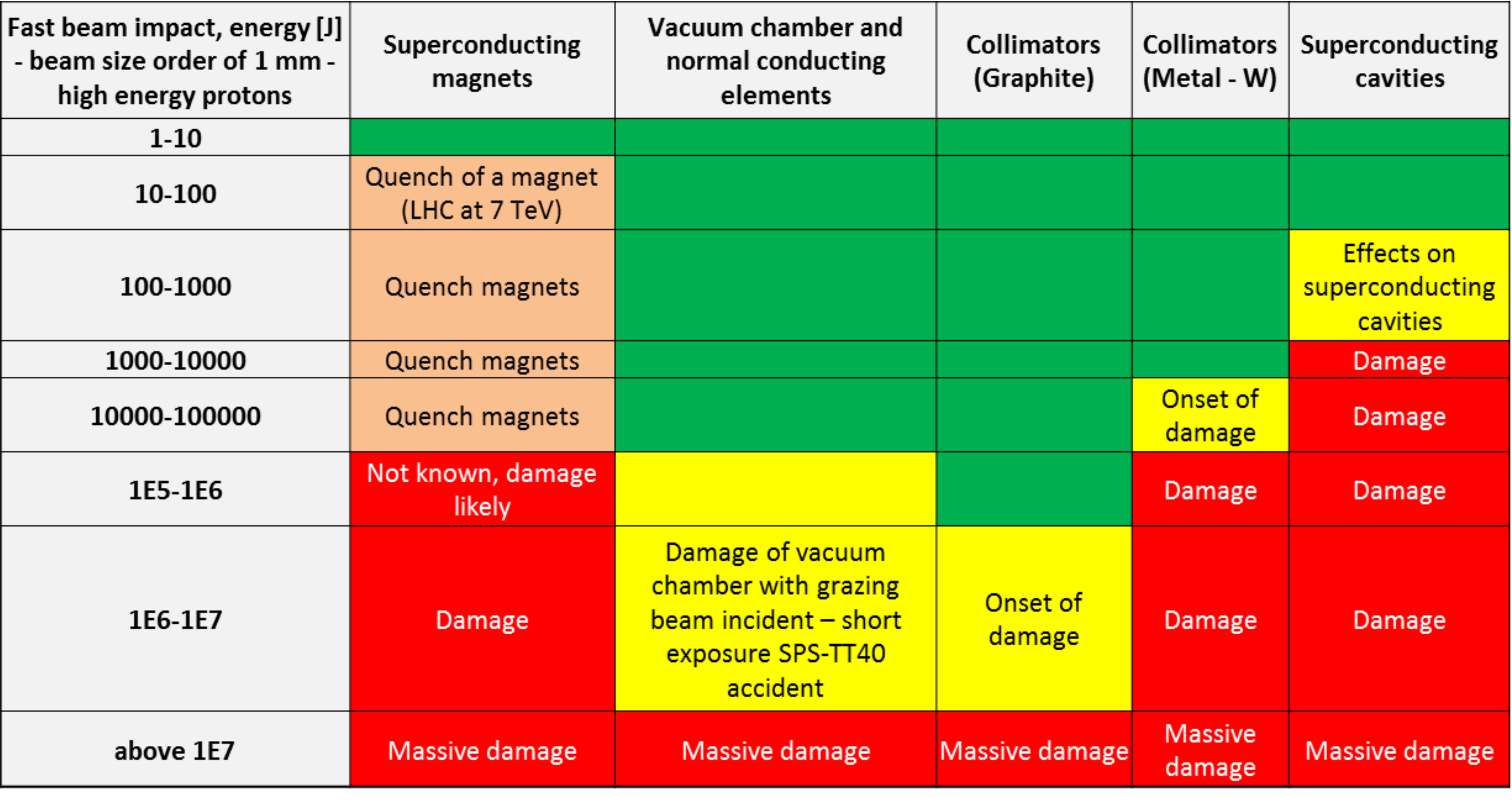}
    \caption{Indication for the quench and damage levels by high-energy protons. Fast beam impact is assumed, with a beam size of the order of 1~mm.}
    \label{damage-table}
\end{figure}

In the following, we discuss the criticality for high-energy proton beams and a beam size in the order of 1 mm. Figure \ref{damage-table} gives an idea of the current knowledge about damage levels. The data in the table is derived from past experience, either from accidents or damage experiments with particle beams. The data should be taken with  care; much more research is required to establish reliable data for different beam and equipment parameters:

\begin{itemize}
	\item TT40 experiment \cite{Kain2014};
	\item SPS damage of UA2 \cite{Schmidt2014};
	\item collimator damage experiments \cite{Redaelli2014};
	\item quench tests at LHC \cite{B.Auchmannetal.2015};
	\item experience from SNS \cite{Plum2014};
	\item TEVATRON accident \cite{Mokhov2014}.
\end{itemize}

When designing protection systems for a specific hazard, a conservative approach is required. However, a too pessimistic approach should be avoided. An accelerator should not be overprotected; this could lead to unnecessary investment and downtime during operation.

\section{Machine protection and interlocks at LHC}

\subsection{Strategy for machine protection}

\begin{figure}[tb]
    \centering
    \includegraphics*[width=0.7\linewidth]{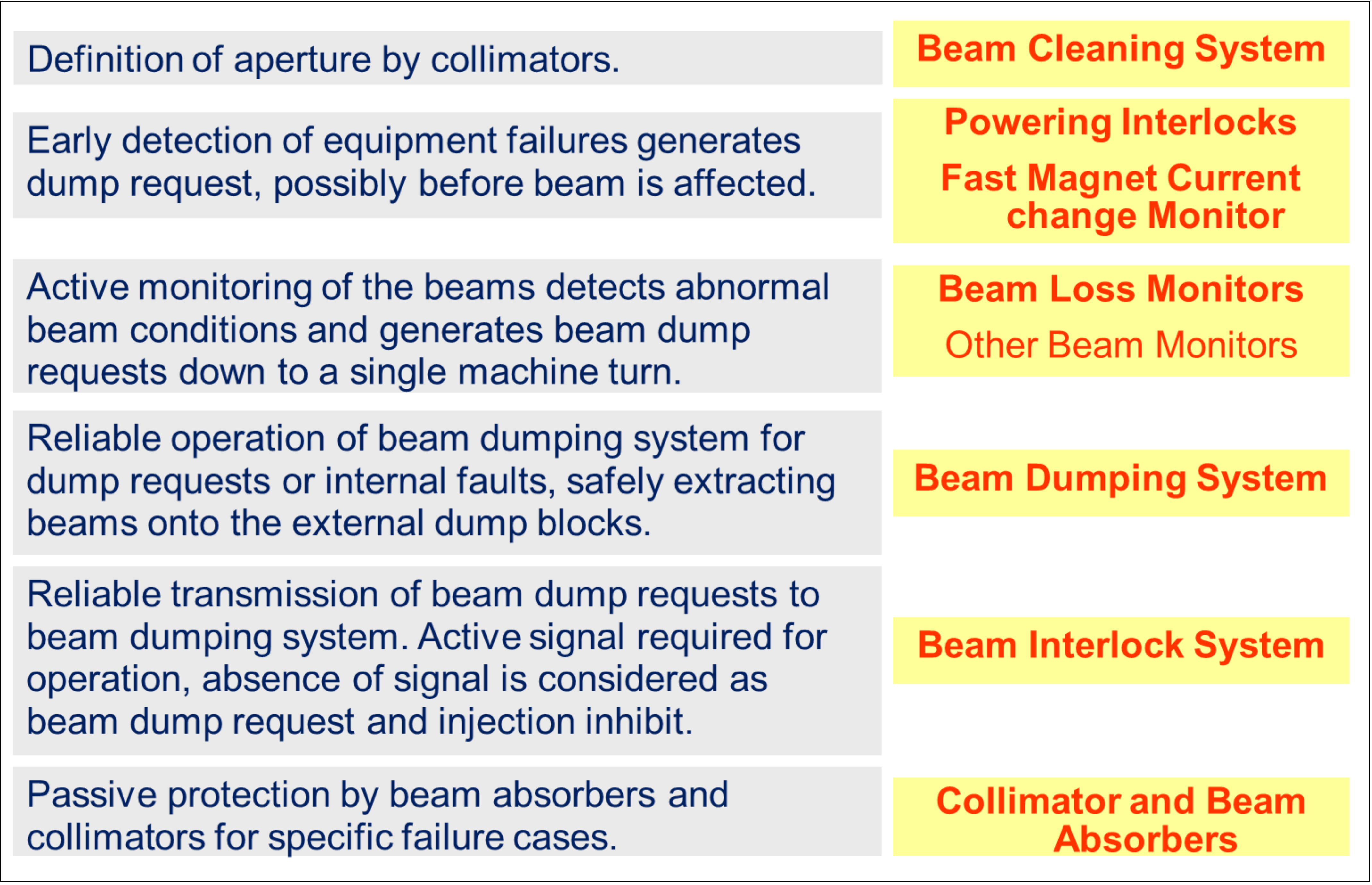}
    \caption{Protection strategy for LHC}
    \label{Protection-Strategy}
\end{figure}

In this section we discuss the strategy adopted for machine protection of LHC and the related systems, illustrated in \Fref{Protection-Strategy}, together with the main subsystems for protection.

\begin{itemize}
	\item Definition of aperture by collimators.
	\item Detect failures at hardware level and stop beam operation for critical failures.
	\item Detect initial consequences of failures on the beam with beam instrumentation before it is too late.
	\item Transmit signal from instrumentation via a highly reliable interlock system to the extraction kickers and injection system.
	\item Stop beam operation by extracting the beams into beam dump block.
	\item Inhibit injection into LHC and extraction from the SPS.
	\item Stop beam by beam absorber/collimator for specific failures.
\end{itemize}

\subsection{Machine interlocks at LHC}

\begin{figure}[tb]
    \centering
    \includegraphics*[width=0.95\linewidth]{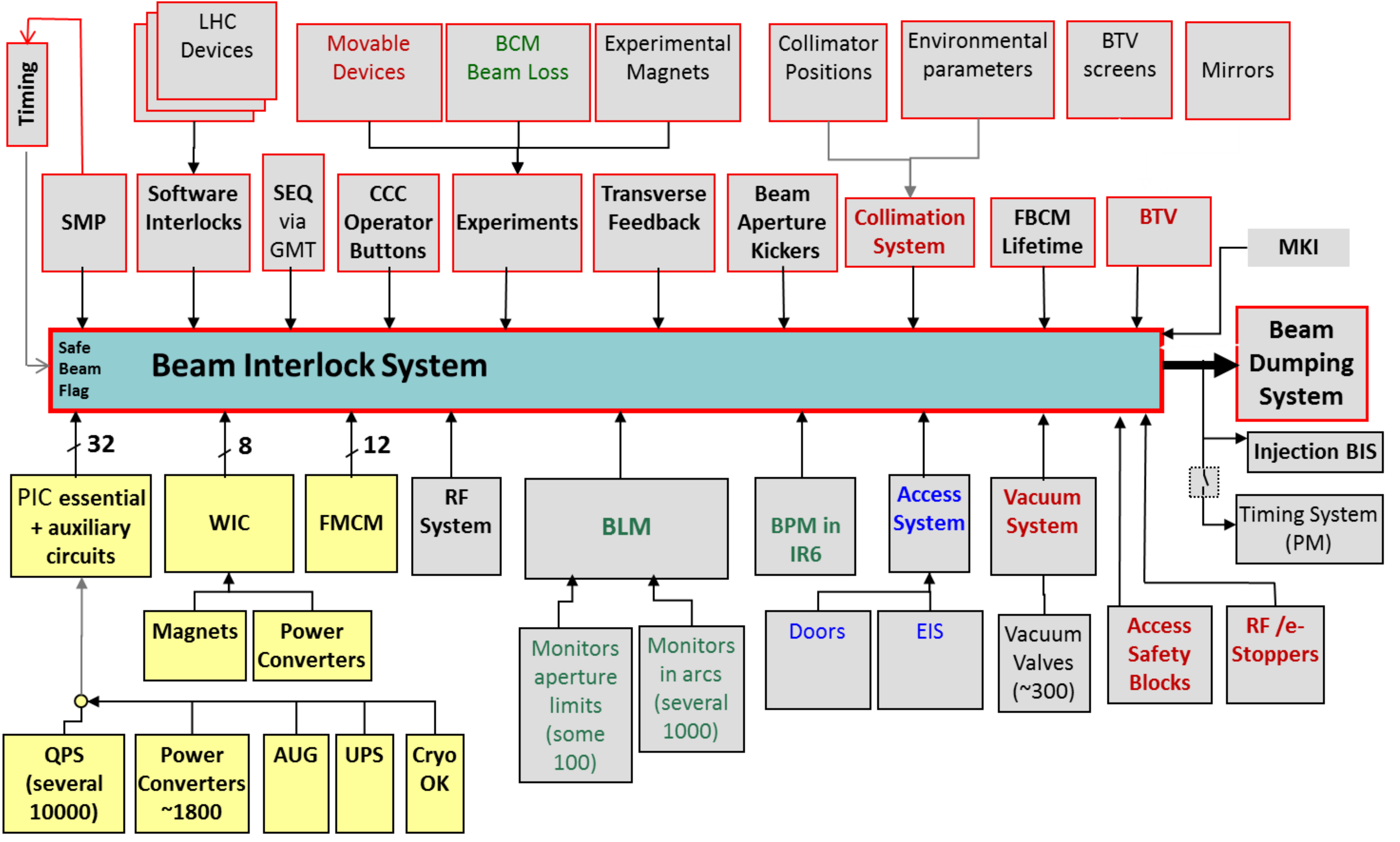}
    \caption{Beam interlock system at LHC as well as connected systems}
    \label{BIS-LHC}
\end{figure}

Figure \ref{BIS-LHC} illustrates the interlock systems for LHC. The heart is the beam interlock system that receives beam dump requests from many connected systems. If a beam dump request arrives, a signal is sent to the beam dumping system to request the extraction of the beams. At the same time, a signal is sent to the injection system to block injection into LHC as well as extraction of beam from the SPS. A third signal is provided for the timing system that sends out a request to many LHC systems for providing data that were recorded before the beam dump, to understand the reasons for the beam dump (typically beam loss, beam position, beam current, magnet currents, etc).

The most complex system of LHC is the superconducting magnets and powering system. The powering interlock system (PIC) ensures communication between systems involved in the powering of the LHC superconducting magnets. This includes power converters, magnet protection system, UPS (un-interruptible power supplies), emergency stop of electrical supplies (AUG) and the cryogenic system. As an example, in case a magnet quench is detected by the quench protection system (QPS), the power converter must stop. In total, there are several tens of thousands of interlock signals. When a failure is detected that risks stopping the powering of magnets, a beam dump request is sent to the beam interlock system. A second system manages interlocks from the normal-conducting magnets and their power supplies (WIC) that ensures protection of normal-conducting magnets in case of overheating.

The machine interlock system is strictly separated from interlocks for personnel safety such as the personnel access system; however, an interlock from the access system is sent to the beam interlock system.

As shown in \Fref{BIS-LHC}, many other systems also provide beam dump requests in case of failure: beam loss monitors, other beam monitors, movable devices and LHC experiments.

\section{Safety and protection integrity levels}

As has been discussed in \cite{Schmidt2014}, the risk for a specific hazard depends on consequences and probability of an accident: ${\rm risk} = {\rm consequences} \times {\rm probability}$.

For the design of protection systems in industry, standards are frequently used. One example of a widely used standard is IEC 61508, an international standard of rules applied in industry, for the Functional Safety of Electrical/Electronic/Programmable Electronic Safety-related Systems (E/E/PE or E/E/PES). For LHC, the standard was not strictly applied, but many ideas and principles were used, such as the safety integrity level (SIL) concept. Since the development of the LHC machine protection system did not strictly follow the IEC 61508 norm, the idea of a protection integrity level (PIL) was introduced \cite{Kwiatkowski2009}.

\begin{figure}[tb]
    \centering
    \includegraphics*[width=0.8\linewidth]{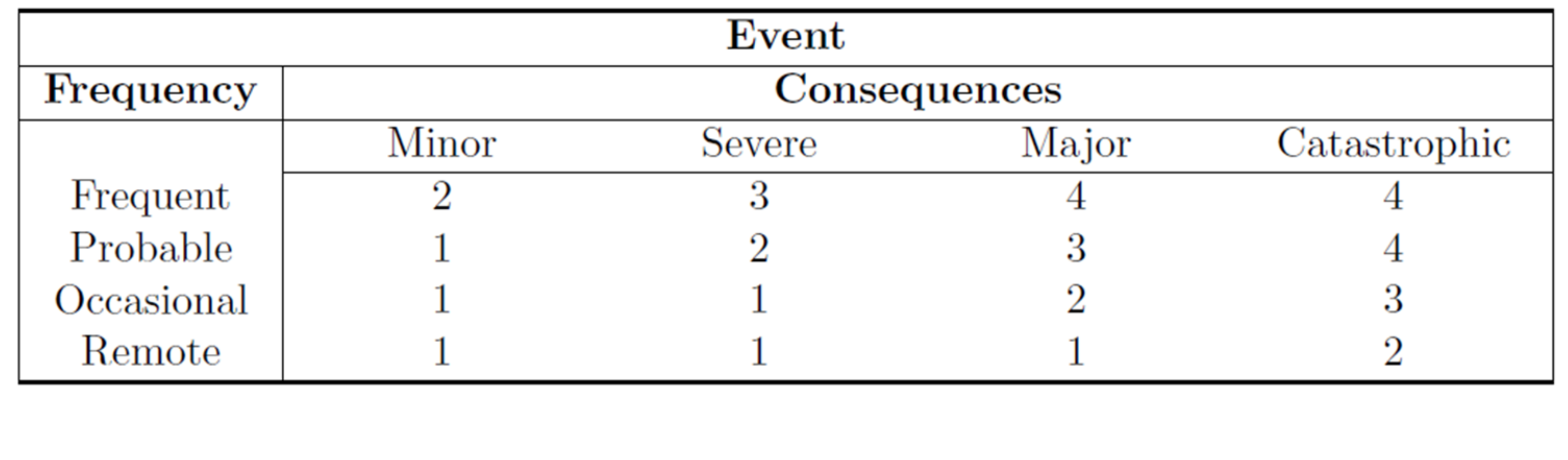}
    \caption{Risk matrix defining the PIL level (protection integrity level), derived from the SIL definition}
    \label{Risk-Matrix}
\end{figure}

The PIL level classifies hazards according to frequency and consequences, as in IEC 61508 for the safety integrity level (see \Fref{Risk-Matrix}). The design of the protection system for a specific hazard depends on the PIL level. The higher the PIL, the more effort is needed to demonstrate that the protection system is sufficiently robust.

\begin{figure}[tb]
    \centering
    \includegraphics*[width=0.8\linewidth]{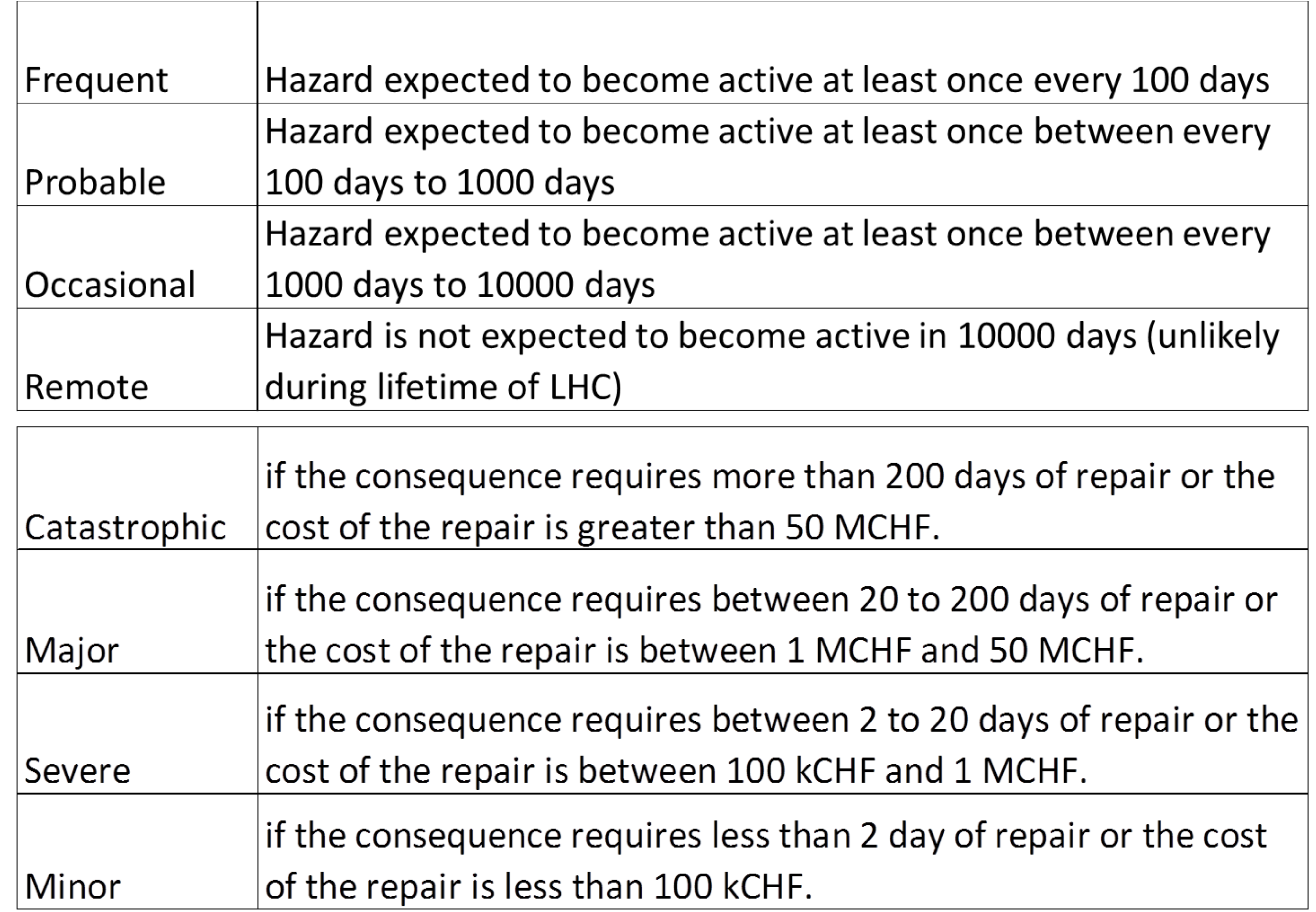}
    \caption{Definition of frequencies and consequences for hazards, defined for LHC}
    \label{SIL-LHC}
\end{figure}

For LHC, a definition for frequency and consequences was adopted as shown in \Fref{SIL-LHC}. The definition of the consequences is not a unique table and here the numbers for LHC are shown that can be very different for other projects.

Some examples of hazards that were analysed during the design of the LHC protection systems are given.

\begin{itemize}
	\item What is the risk when a probe (= low-intensity) beam is injected into the empty LHC ring and lost during the first turn? Low risk since the consequences of such event are minor: PIL 1.
	\item What is the risk that a power converter fails and the full LHC beam is deflected? High risk since the probability for such an event is high (probable) and without protection the consequences would be major or catastrophic for LHC: PIL 3 or PIL 4.
	\item What is the risk for a wire scanner accidentally moving through the high-intensity beam? The risk is low since the consequences are minor and the frequency occasional: PIL 1.
	\item What is the risk that the entire LHC beam will be deflected by a black hole generated in a proton--proton collision  in ATLAS? Remote probability, not to be considered.
\end{itemize}

\subsection{Design considerations for protection systems}

There are a number of principles that should be considered in the design of protection systems, although it might not be possible to follow all these principles in all cases.

\begin{itemize}

  \item Fail-safe design: in case of a failure in the protection system, protection functionalities should not be compromised. As an example, if the cable from the interlock system that triggers the extraction kicker of the beam dumping system is disconnected, the kicker must fire and operation must stop.

  \item Detection of internal faults: the protection system must monitor its internal status. In case of an internal fault, the fault should be reported. If the fault is critical, operation must be stopped.

  \item Remote testing should be an integral part of the design, for example between two runs. This allows regular verification of the correct status of the system.

  \item Critical equipment should be redundant (possibly diverse redundancy, with the same or similar functions executed by different systems).

  \item Critical processes for protection should not rely on complex software running under an operating system and requiring the general computer network.

  \item It should not be possible to remotely change the most critical parameters. If parameters need to be changed, the changes must be controlled, logged and password protection should ensure that only authorized personnel can perform the change.

  \item Safety, availability and reliability of the systems should be demonstrated. This is possible by using established methods to analyse critical systems and to predict failure rates.

  \item Operate the protection systems early on before they become critical, to gain experience and to build up confidence. This could be done before beam operation or during early beam operation when the beam intensity is low.

  \item It is inevitable to disable interlocks (e.g. during early phase of commissioning and for specific tests). Managing interlocks (e.g. disabling) is common practice and required in the early phase of operation. Keep track and consider masking of interlocks during the system design. Example for the realization at LHC: masking of some interlocks is possible, but only for low-intensity/low-energy beams (`safe beams').

  \item Avoid (unnecessary) complexity for protection systems, keep it simple.

  \item Having a vision to the operational phase of the system helps to defined the functionality.

  \item Test benches for electronic systems should be part of the system development.

  \item Most failures in electronics systems are due to power supplies, cables, mechanical parts and connectors. Redundancy of these parts will increase the availability.

  \item Careful testing in conditions similar to real operation is required.

  \item Reliable protection does not end with the development phase. Documentation for installation, commissioning, maintenance and operation of the MPS is required.

  \item The accurate execution of each protection function must be explicitly tested during commissioning.

  \item Requirements need to be established for the test interval of each function.

   \item All actions of the protection systems (e.g. beam dumps) need to be carefully analysed. This requires the presence of transient recording of all relevant systems.

\end{itemize}

\section{Summary}

In systems engineering, dependability is a measure of a system's availability, reliability and maintainability. This concept is a new challenge for accelerator laboratories and  requires a different approach in engineering, operation and management.

Safety culture is how safety is managed in the laboratory and  reflects attitudes, beliefs, perceptions and values that employees share in relation to safety. Safety culture at CERN has been developed over the last, say, 10 years. Even those who were sceptical about the need for a powerful protection system were convinced after the accident during the powering test in 2008 that such safety culture is needed.

At CERN for LHC, the experience with the systems for protecting equipment from beam accidents is excellent; there was no damage and no near miss. However, some non-conformities were detected that demonstrate that it is important to be vigilant.

Availability and safety are a trade-off relationship---when a given safety is met, the goal is to make the system as available as possible for providing beams to experiments. False beam dumps (beam dumps that are not strictly necessary for safe operation) need to be avoided to minimize downtime. The availability at LHC is not yet at a level that is acceptable for future operation and improvements are required.

From the experience with LHC there are a number of lessons to be learned for future accelerators to ensure safe operation with high availability (e.g. for accelerator-driven spallation, where operating with very high availability is the main challenge).

Machine protection is not equal to equipment protection and is not limited to the interlock system:
\begin{itemize}
 \item it requires a thorough understanding of many different types of failures that could lead to beam loss;
 \item it requires a comprehensive understanding of all aspects of the accelerator (accelerator physics, operation, equipment, instrumentation, functional safety);
 \item it touches many aspects of accelerator construction and operation and includes many systems;
 \item it is becoming increasingly important for future projects, with increased beam power/energy density (W/mm$^{2}$ or J/mm$^{2}$) and increasingly complex machines.
 \end{itemize}

\section*{Acknowledgements}

I wish to thank many colleagues from CERN, ESS and the authors of the listed papers for their help and for providing material for this paper.

%\section{Bibliography}
%\bibliographystyle{ieeetr}
%\bibliography{\myreferences/Rudi-bibliography,\myreferences/Tahir,\myreferences/SPS-bibliography,\myreferences/Other-papers}

\begin{thebibliography}{10}

\bibitem{TheLHCStudyGroup1995}
The LHC Study Group, The Large Hadron Collider: conceptual design,
CERN/AC/95-05 (LHC) (1995).

\bibitem{Kain2014}
V.~Kain, Beam transfer and machine protection, these proceedings.

\bibitem{Kain2014a}
V.~Kain, Beam losses in circular accelerators, these proceedings.

\bibitem{Plum2014}
M.~Plum, Beam dynamics and beam losses -- linear machines, these proceedings.

\bibitem{Redaelli2014}
S.~Redaelli, Beam cleaning and collimation systems, these proceedings.

\bibitem{Yee-Rendon2014}
B.~Yee-Rendon, R.~Lopez-Fernandez, R.~Calaga, R.~Tomas, F.~Zimmermann and
  J.~Barranco, Fast crab cavity failures in HL-LHC, 5th
  Int. Particle Accelerator Conf., IPAC 2014, Dresden, Germany,
  15--20 June 2014.

\bibitem{Gomez-Alonso2009}
A.~Gomez-Alonso, Ph.D. thesis, UPC Barcelona, CERN-THESIS-2009-02, 2009.

\bibitem{Werner2005}
M.~Werner, M.~Zerlauth, R.R. Schmidt, V.~Kain and B.~Goddard, A fast magnet
  current change monitor for machine protection in HERA and the LHC,
  ICALEPCS 2005, Geneva, Switzerland, 2005.

\bibitem{Dehning2014}
B.~Dehning, Beam loss monitors at LHC, these proceedings.

\bibitem{Tahir2012}
N.A. Tahir, J.B. Sancho, A.~Shutov, R.~Schmidt and A.R. Piriz, {\em Phys. Rev. ST Accel. Beams} \textbf{15} (2012) 051003. http://dx.doi.org/10.1103/PhysRevSTAB.15.051003

\bibitem{fluka}
G. Battistone \emph{et al.}, {\em AIP Conf. Proc.} \textbf{896} (2007) 31. http://dx.doi.org/10.1063/1.2720455

\bibitem{Wilson-1993-PAC}
D.C. Wilson \emph{et al}., Hydrodynamic calculations of 20-TeV beam interactions
  with the SSC beam dump, Proc. PAC 1993, San Sebastian,
  Spain, 1993. http://dx.doi.org/10.1109/pac.1993.309562

\bibitem{Blanco-HB2012}
J. Blanco \emph{et al}., An experiment on hydrodynamic tunneling of the SPS high
  intensity proton beam at the HIRdmat facility, Proc.
  HB2012, Beijing, China, 2012.

\bibitem{Schmidt2014}
R.~Schmidt, Introduction to machine protection, these proceedings.

\bibitem{B.Auchmannetal.2015}
B. Auchmann \emph{et al}., Testing beam-induced quench levels of LHC
  superconducting magnets in run 1, submitted to {\em Phys. Rev. Spec. Top.
  Accel. Beams}, June 2015.

\bibitem{Mokhov2014}
N.~Mokhov, Beam material interaction, heating and activation, these proceedings.

\bibitem{Kwiatkowski2009}
M.~Kwiatkowski, Ph.D. thesis, Warsaw University of Technology, Faculty of Electronics and Information Technology, CERN-THESIS-2014-048, 2009.

\end{thebibliography}

\end{document}